\documentclass[final,3p,times]{elsarticle}

\usepackage{amssymb}
\usepackage{graphicx,setspace}
\usepackage{psfrag}
\usepackage{amsfonts}
\expandafter\let\csname equation*\endcsname\relax
\expandafter\let\csname endequation*\endcsname\relax
\usepackage{amsmath}
\usepackage{amssymb, amsthm}
\usepackage{mathrsfs}
\usepackage{epstopdf}
\usepackage{float}
\usepackage{lineno,hyperref}
\usepackage{color}
\usepackage{subfigure}
\usepackage{amsmath}
\usepackage{dsfont} 
\usepackage{amssymb} 
\usepackage{graphicx,color}
\usepackage{extarrows}

\usepackage{graphicx}

\usepackage{epstopdf}

\journal{ArXiv}

\begin{document}
\newtheorem{definition}{Definition}[section]
\newtheorem{lemma}{Lemma}[section]
\newtheorem{remark}{Remark}[section]
\newtheorem{theorem}{Theorem}[section]
\newtheorem{proposition}{Proposition}
\newtheorem{assumption}{Assumption}
\newtheorem{example}{Example}
\newtheorem{corollary}{Corollary}[section]
\def\ep{\varepsilon}
\def\Rn{\mathbb{R}^{n}}
\def\Rm{\mathbb{R}^{m}}
\def\E{\mathbb{E}}
\def\hte{\hat\theta}
\renewcommand{\theequation}{\thesection.\arabic{equation}}
\begin{frontmatter}



\title{Dynamics of Coupled Stochastic van der Pol Oscillators: Bifurcations, Synchronization and Chaos}

\author{Shenglan Yuan\fnref{addr1}}\ead{shenglanyuan@gbu.edu.cn}
\author{Xiang Zhou\corref{cor1}\fnref{addr2}}\ead{xizhou@cityu.edu.hk}\cortext[cor1]{Corresponding author}
\address[addr1]{\rm Department of Mathematics, School of Sciences, Great Bay University, Dongguan 523000, China }
\address[addr2]{\rm Department of Mathematics, City University of Hong Kong, Kowloon, Hong Kong }

\begin{abstract}
This work presents a comprehensive analysis of coupled stochastic van der Pol oscillators, a paradigm for understanding synchronization, bifurcations, and chaos in nonlinear systems subject to random fluctuations. The system comprises two or more oscillators with nonlinear damping, linear diffusive coupling, and additive Gaussian white noise.
We develop a unified framework that systematically connects global bifurcations, synchronization phenomena, and chaotic dynamics within a single coherent stochastic model. We explore the stochastic dynamics of coupled van der Pol oscillators by seamlessly blending theoretical principles with in-depth numerical simulations. This integrated approach forms a robust framework for analysis, with essential phenomena clearly depicted in the accompanying figures. We then extend this framework to a comprehensive investigation of large networks, focusing on their continuum limit, emergent pattern formation, the role of noise, and the onset of collective chaos.
\end{abstract}

\end{frontmatter}

\section{Introduction}
The system traces its roots to the classical work of Balthasar van der Pol in the 1920s, who pioneered the study of relaxation oscillations in electrical circuits (triode generators). The equation he developed became a fundamental model for self-sustained oscillations \cite{BVdP}. The study of coupled oscillators evolved from this foundation. The specific model of two coupled van der Pol oscillators is a natural extension for studying interaction phenomena, further popularized by Winfree and Kuramoto in the 1970s, who shifted focus to the collective dynamics of large populations \cite{ATW,YK}. Storti and Rand \cite{SR} investigated the dynamics of two strongly coupled van der Pol oscillators. Early analysis focused on coupled van der Pol oscillators and synchronization transitions.

The coupled stochastic van der Pol oscillators system
\begin{equation}\label{SVdP}
\left\{
  \begin{array}{ll}
    \ddot{x}_1-\mu(1-x_1^{2})\dot{x}_1+\omega_1^{2}x_1=\kappa(x_2-x_1)+\sigma_1\xi_1(t), &  \\
    \ddot{x}_2-\mu(1-x_2^{2})\dot{x}_2+\omega_2^{2}x_2=\kappa(x_1-x_2)+\sigma_2\xi_2(t), &
  \end{array}
\right.
\end{equation}
comprises two coupled van der Pol oscillators with independent or correlated noise. This system represents a modern synthesis of classical ideas, incorporating nonlinear damping $\mu(1-x_i^2)\dot{x}_i$ (the hallmark of the van der Pol oscillator), mutual coupling $\kappa(x_j - x_i)$ (linear diffusive coupling), and stochastic forcing $\sigma_i\xi_i(t)$ (additive Gaussian white noise). We are interested in how noise and coupling affect the dynamics, particularly global bifurcations (changes in the structure of attractors), synchronization (phase locking and frequency locking),
and chaos (deterministic and noise-induced).

The system \eqref{SVdP} is not merely theoretical, it has significant real-world applications across multiple disciplines \cite{ES,YW}. Rings of coupled van der Pol oscillators under stochastic forcing model thermo- and aeroacoustic interactions of sound fields in can-annular combustors, helping to  understanding damaging high-amplitude pressure oscillations that lead to high-cycle fatigue in engine components \cite{PR}. The principle of stochastic resonance is exploited in engineering, where coupled van der Pol and Duffing oscillator systems utilize noise to detect characteristic frequencies, with applications such as bearing fault diagnosis. The original van der Pol oscillator was an electrical circuit model; coupled versions model arrays of interacting electronic oscillators \cite{MD}. The study of oscillator populations is crucial for modeling systems like spiking neurons and central pattern generators; understanding synchronization \cite{PRK} and  its control has direct implications for medical treatments of neurological disorders such as Parkinson's disease \cite{ET}.
Coupled oscillators model pacemaker cells and the propagation of electrical signals in the heart. Simplified models of the El Ni\~{n}o-Southern Oscillation often use relaxation oscillators, and coupling such oscillators can represent teleconnections between different oceanic basins \cite{JG}.

Afraimovich \emph{et al}. \cite{VA} explored the nonlinear dynamics of specialized discrete media composed of interconnected phase synchronization systems.
Acebr\'on \emph{et al}. \cite{JAA} investigated the continuum models for synchronization in coupled oscillators. Goldobin and Pikovsky \cite{GP} analyzed common noise effects on the synchronization and desynchronization dynamics of self-sustained oscillators, including phase locking.
Lindner \emph{et al}. \cite{BL} explored the effects of noise in excitable systems, including the van der Pol oscillator in the excitable regime. Nakao \cite{HN} studied phase reduction methods for stochastic nonlinear oscillators.
Nagai and Kori \cite{NK} analyzed noise-induced synchronization in a large population of globally coupled nonidentical oscillators, focusing on dynamics with heterogeneous frequencies and noise. P\'erez-Cervera, Lindner, and Thomas \cite{PCLT} applied phase reduction to stochastic oscillators via isostable-based methods. Teramae and Tanaka \cite{TT} examined the robustness of phase synchronization induced by common noise in a broad class of limit cycle oscillators.

Unlike many studies that focus on a single aspect (e.g., only synchronization or only bifurcations), this work provides a unified analysis of global bifurcations, synchronization, and chaos within a single coherent stochastic model \eqref{SVdP}. It systematically connects these phenomena, showing how they interact under the influence of noise. The significance lies in its synthesis and depth. It takes a classical model and provides a state-of-the-art analysis in the context of modern stochastic dynamics. By systematically exploring the interplay between nonlinearity, coupling, and the correlation structure of noise, we offer a powerful framework for understanding complex phenomena in any system that can be modeled by coupled self-sustained oscillators.

For the deterministic case, we analyze symmetry properties, fixed point stability, and the emergence of limit cycles, establishing conditions for phase and frequency locking in both identical and non-identical oscillators. The bifurcation structure is characterized, including saddle-node on invariant circle (SNIC) and homoclinic bifurcations that govern transitions between synchronized and desynchronized regimes.

In the stochastic setting, we investigate how noise correlation structure fundamentally alters system behavior. Through stochastic averaging and phase reduction, we derive effective Langevin equations for the phase difference and quantify synchronization via phase coherence measures. We demonstrate that independent noise disrupts synchronization through phase diffusion and occasional phase slips, while common noise paradoxically enhances synchronization through noise-induced contraction mechanisms and effective frequency mismatch reduction. The analysis encompasses stochastic P-bifurcations (changes in stationary probability density topology) and D-bifurcations (changes in Lyapunov exponent sign \cite{PP}), revealing how noise smooths deterministic transitions while introducing new phenomena such as noise-induced limit cycles and stochastic resonance.

Chaos in the stochastic system \eqref{SVdP} is characterized through Lyapunov exponents, Kolmogorov-Sinai entropy, and Kaplan-Yorke dimension. We employ stochastic Melnikov analysis to predict noise-induced homoclinic chaos and demonstrate that even deterministically regular systems can exhibit positive Lyapunov exponents when noise interacts with near-bifurcation dynamics.

Extensive numerical simulations using Euler-Maruyama integration validate theoretical predictions across diverse parameter regimes \cite{DJH}. We systematically map synchronization transitions as functions of coupling strength, frequency mismatch, and noise intensity, identifying Arnold tongues and regions of noise-enhanced synchronization. Phase slip analysis reveals Kramers-type escape rates over effective potential barriers, with inter-slip intervals following Poisson statistics. Global bifurcation detection via Poincar\'e maps captures period divergence at homoclinic bifurcations, with period scaling logarithmically near critical coupling values.

Extending to large networks, we formulate continuum limit descriptions as stochastic partial differential equations, analyzing pattern formation, spatial correlation functions, and structure factors. The dispersion relation reveals how noise excites spatial modes absent in deterministic dynamics, leading to noise-induced Turing patterns even with positive diffusion coefficients. Collective chaos emerges in globally coupled ensembles, characterized by order parameter fluctuations and finite-size effects.

In Section \ref{DC}, we exposit the deterministic dynamics of two coupled van der Pol oscillators. It covers the system's formulation, symmetry, fixed points, limit cycles, synchronization, and bifurcations. In Section \ref{SC}, we investigate the stochastic dynamics of coupled van der Pol oscillators. This section expertly blends theoretical concepts with detailed descriptions of numerical simulations, creating a robust framework for analysis. We present the essential aspects of stochastic synchronization, bifurcations, and chaos, and the figures are designed to illustrate these phenomena effectively. In Section \ref{LN}, we provide a comprehensive study of large networks of coupled van der Pol oscillators, focusing on their continuum limit, pattern formation, noise-induced phenomena, and collective chaos. In Section \ref{CFC}, we summarize our findings, present our conclusions, and outline directions for future research.

\section{Deterministic Case}\label{DC}
First, we consider the deterministic system ($\sigma_1=\sigma_2=0$). The equations are
\begin{equation*}
\left\{
  \begin{array}{ll}
    \ddot{x}_1-\mu(1-x_1^{2})\dot{x}_1+\omega_1^{2}x_1=\kappa(x_2-x_1), &  \\
    \ddot{x}_2-\mu(1-x_2^{2})\dot{x}_2+\omega_2^{2}x_2=\kappa(x_1-x_2). &
  \end{array}
\right.
\end{equation*}
This system describes two coupled van der Pol oscillators with natural frequencies $\omega_1$ and $\omega_2$, damping parameter $\mu$, and coupling strength $\kappa$.
We can rewrite it as a system of four-dimensional first-order ordinary differential equations. Let $y_1=\dot{x}_1$, $y_2=\dot{x}_2$. Then
\begin{equation}\label{VdP}
\left\{
\begin{array}{ll}
\dot{x}_1=y_1, &  \\
\dot{y}_1=\mu(1-x_1^{2})y_1-\omega_1^{2}x_1+\kappa(x_2-x_1), & \\
\dot{x}_2=y_2, & \\
\dot{y}_2=\mu(1-x_2^{2})y_2-\omega_2^{2}x_2+\kappa(x_1-x_2). &
\end{array}
\right.
\end{equation}
\subsection{Symmetry and fixed points}
The trivial equilibrium $(x_1,y_1,x_2,y_2)=(0,0,0,0)$ of system \eqref{VdP} exists for all parameters. Linearizing around the origin gives
\begin{equation*}
\left\{
\begin{array}{ll}
\dot{x}_1=y_1, &  \\
\dot{y}_1=\mu y_1-\omega_1^{2}x_1+\kappa(x_2-x_1), & \\
\dot{x}_2=y_2, & \\
\dot{y}_2=\mu y_2-\omega_2^{2}x_2+\kappa(x_1-x_2). &
\end{array}
\right.
\end{equation*}
The Jacobian at the origin is
\begin{equation*}
J=\left(
    \begin{array}{cccc}
      0 & 1 & 0 & 0 \\
      -\omega_1^{2}-\kappa & \mu & \kappa & 0 \\
      0 & 0 & 0 & 1 \\
      \kappa & 0 & -\omega_2^{2}-\kappa & \mu \\
    \end{array}
  \right).
\end{equation*}
The characteristic equation is
\begin{equation*}
\det(J-\lambda I)=0.
\end{equation*}
This is a fourth-degree polynomial. For $\mu>0$, the origin is unstable (the van der Pol oscillator has an unstable fixed point and a stable limit cycle).

The system \eqref{VdP} has a symmetry exchanging oscillator 1 and 2 when $\omega_1=\omega_2=\omega_0$.
For identical oscillators, the characteristic polynomial factors are described by
\begin{equation}\label{ch}
(\lambda^2-\mu\lambda+\omega_0^2)(\lambda^2-\mu\lambda+\omega_0^2+2\kappa)=0.
\end{equation}

\subsection{Limit cycles and synchronization}
For uncoupled oscillators ($\kappa=0$), each reduces to the classic van der Pol oscillator:
\begin{equation*}
 \ddot{x}_i-\mu(1-x_i^{2})\dot{x}_i+\omega_i^{2}x_i=0,\quad i=1,2.
\end{equation*}
When $\mu>0$, each oscillator exhibits a stable limit cycle due to the balance between nonlinear damping $-\mu x_i^{2}\dot{x}_i$ and negative linear damping $\mu\dot{x}_i$.

With coupling ($\kappa>0$), the oscillators may synchronize in frequency and phase. We can observe phase locking (the phases of the two oscillators maintain a constant difference) and frequency locking (the frequencies become equal).

For identical oscillators ($\omega_1=\omega_2=\omega_0$), the system has symmetry
$(x_1,y_1)\leftrightarrow(x_2,y_2)$. Synchronized solutions satisfy
\begin{equation*}
x_1(t)=x_2(t),\quad y_1(t)=y_2(t),
\end{equation*}
reducing to a single van der Pol oscillator.
The synchronization manifold $\{x_1=x_2, y_1=y_2\}$ is invariant. The stability of the synchronized state can be analyzed by considering the transverse modes (difference between the two oscillators). Define difference variables $\delta x=x_1-x_2$, $\delta y=y_1-y_2$.
The linearized transverse dynamics yield
\begin{equation*}
\left\{
\begin{array}{ll}
\delta\dot{x}=\delta y, & \\
\delta\dot{y}=\mu(1-\bar{x}^2)\delta y-2\mu\bar{x}\bar{y}\delta x-(\omega_0^2+2\kappa)\delta x, &
\end{array}
\right.
\end{equation*}
where $\bar{x}$ and $\bar{y}$ are coordinates on the synchronization manifold.
The transverse Jacobian (for the difference variable) can be computed and its eigenvalues determine the stability.

The eigenvalues are
\begin{equation*}
\lambda=\frac{\mu\pm\sqrt{\mu^2-4(\omega_0^2+2\kappa)}}{2}.
\end{equation*}
For $\kappa>-\omega_0^2/2$ (always true if $\omega_0^2>0$), the synchronized state is stable if all transverse eigenvalues have negative real parts. For $\mu>0$, stability requires additional conditions (typically holds for sufficiently strong coupling $\kappa$).

For non-identical oscillators ($\omega_1\neq\omega_2$), phase reduction (for weak coupling and small $\mu$) leads to a phase-difference equation. Let $\phi_1$, $\phi_2$ be phases. Then
\begin{align*}
\dot{\phi}_1&=\omega_1+\kappa\Gamma(\phi_2-\phi_1), \\
\dot{\phi}_2&=\omega_2+\kappa\Gamma(\phi_1-\phi_2).
\end{align*}
Defining $\phi=\phi_1-\phi_2$,
\begin{equation*}
\dot{\phi}=\Delta\omega+\kappa\big(\Gamma(-\phi)-\Gamma(\phi)\big),
\end{equation*}
where $\Delta\omega=\omega_1-\omega_2$. The function $\Gamma$ is the coupling function derived via averaging. Fixed points $\phi^{\ast}$ correspond to phase-locked states. Synchronization occurs if
\begin{equation*}
|\Delta \omega|\leq\kappa|\max\Gamma_{\text{diff}}|,
\end{equation*}
where $\Gamma_{\text{diff}}(\phi)=\Gamma(-\phi)-\Gamma(\phi)$.
The critical coupling for synchronization is
\begin{equation*}
\kappa_{c}\sim\frac{\Delta\omega^2}{2\omega_0}\quad (\text {for weak coupling}).
\end{equation*}
\subsection{Bifurcations}
As the parameters $(\kappa,\mu,\omega_1,\omega_2)$ vary, the system \eqref{VdP} can undergo saddle-node bifurcations of limit cycles, Hopf bifurcations (for the synchronized state or other periodic orbits), and global bifurcations such as homoclinic bifurcations.

For identical oscillators, the symmetric synchronized state can lose stability in a symmetry-breaking bifurcation, leading to antisymmetric or out-of-phase oscillations.
Base on the characteristic polynomial factors in \eqref{ch}, there is a Hopf bifurcation at $\mu=0$. As $\mu$ increases from negative to positive, two pairs of complex conjugate eigenvalues cross the imaginary axis simultaneously (provided $\omega_0^2>0$ and
$\omega_0^2+2\kappa>0$). This is a double Hopf bifurcation, generating two limit cycles (in-phase and anti-phase modes).

In particular, when the oscillators are non-identical ($\omega_1\neq\omega_2$), the system may exhibit a bifurcation from a synchronized state to a desynchronized state as the coupling strength $\kappa$ changes. This can be a saddle-node on invariant circle  bifurcation or a Hopf bifurcation leading to a torus and then to chaos.

A saddle-node on invariant circle bifurcation occurs when a limit cycle collides with  a saddle, creating a homoclinic orbit. In coupled system, this corresponds to the transition between phase-locked and drifting regimes.

For $\mu\gg1$ (relaxation oscillations), the synchronized state can undergo a homoclinic bifurcation, which creates bursting behavior when perturbed.

A SNIC bifurcation requires a saddle-node bifurcation occurring on a closed invariant curve. In this system, it can arise in the phase-reduced model for the non-identical case. Consider the phase-difference equation
\begin{equation*}
\dot{\phi}=\Delta \omega+\kappa\Gamma_{\text{diff}}(\phi),
\end{equation*}
where $\Gamma_{\text{diff}}(\phi)$ is an odd periodic function (e.g., $\Gamma_{\text{diff}}(\phi)=-2\sin(\phi)$ for weak nonlinearity). A SNIC occurs when $\Delta \omega\approx \kappa|\max\Gamma_{\text{diff}}|$, causing a saddle-node bifurcation that creates a phase-locked state on the invariant circle (the torus). At the bifurcation, the saddle and node collide, leaving a homoclinic connection to the former saddle, which becomes a global invariant circle.

In the $(\kappa,\Delta\omega)$ parameter space, the phase-locked region is defined by $\{(\kappa,\Delta\omega): |\Delta\omega|<\kappa f(\mu)\}$, where the function $f(\mu)$ depends on the oscillation waveform: for nearly sinusoidal oscillators ($\mu\ll1$), $f(\mu)$ approaches a constant, whereas for relaxation oscillators ($\mu\gg1$), it scales as $f(\mu)\sim1/\mu$.

\section{Stochastic Case}\label{SC}
Now we add noise. The coupled stochastic van der Pol oscillator system \eqref{SVdP} has the state-space representation
\begin{equation}\label{CSVdP}
\left\{
\begin{array}{ll}
dx_1=y_1dt, &  \\
dy_1=[\mu(1-x_1^{2})y_1-\omega_1^{2}x_1+\kappa(x_2-x_1)]dt+\sigma_1dW_1, & \\
dx_2=y_2dt, & \\
dy_2=[\mu(1-x_2^{2})y_2-\omega_2^{2}x_2+\kappa(x_1-x_2)]dt+\sigma_2dW_2, &
\end{array}
\right.
\end{equation}
where $W_1$ and $W_2$ are independent Wiener processes. The noise correlation structure can be independent ($\langle dW_1 dW_2 \rangle = 0$), common ($dW_1 = dW_2 = dW$), or partially correlated ($\langle dW_1 dW_2 \rangle = \rho dt$).

\subsection{Stochastic synchronization}
Noise can induce various forms of synchronization even when deterministic coupling fails.
We consider identical oscillators with $\omega_1=\omega_2=\omega_0$. Define $u=x_1-x_2$, $v=y_1-y_2$. Subtracting the equations in \eqref{CSVdP} yields
\begin{align*}
du&=vdt, \\
dv&=[\mu v-(\omega_0^{2}+2\kappa)u-\mu(x_1^2y_1-x_2^2y_2)]dt+\sigma_1dW_1-\sigma_2dW_2.
\end{align*}
The nonlinear term complicates exact analysis. For weak coupling and near synchronization, linearize around $u=v=0$ gives
\begin{equation*}
d\left(
   \begin{array}{c}
     u \\
     v \\
   \end{array}
 \right)=\underbrace{\left(
           \begin{array}{cc}
             0 & 1 \\
             -(\omega^2+2\kappa) & \mu \\
           \end{array}
         \right)}_{A}\left(
   \begin{array}{c}
     u \\
     v \\
   \end{array}
 \right)dt+\left(
             \begin{array}{c}
               0 \\
               \sigma_1dW_1-\sigma_2dW_2 \\
             \end{array}
           \right).
\end{equation*}

The linearized difference system is a two-dimensional Ornstein-Uhlenbeck process. Its mean-square stability is determined by the eigenvalues of $A$. If all eigenvalues have negative real parts, the difference decays in mean square. However, for $\mu>0$, the trace $\mu>0$, so at least one eigenvalue has a positive real part, making the synchronized state unstable in the deterministic sense. Noise can nevertheless induce phase synchronization.

The phase coherence, defined as
\begin{equation*}
R_{\phi}(t)=|\langle e^{i[\phi_1(t)-\phi_2(t)]}\rangle|,
\end{equation*}
serves as a measure of synchronization, where the phases are extracted via the Hilbert transform.
The synchronization error is quantified by
\begin{equation*}
E_{\text{sync}}=\langle(x_1-x_2)^2+(y_1-y_2)^2\rangle,
\end{equation*}
while the Lyapunov exponent for transverse directions is defined as
\begin{equation*}
\lambda_{\perp}=\lim_{t\rightarrow\infty}\frac{1}{t}\ln\frac{\|\delta_{\textbf{x}_{\perp}}(t)\|}{\|\delta_{\textbf{x}_{\perp}}(0)\|},
\end{equation*}
with the synchronization condition being $\lambda_{\perp} < 0$ (almost surely or in mean).

For identical oscillators and equal noise intensities $(\sigma_1=\sigma_2)$, the transverse Lyapunov exponent $\lambda_{\perp}$ determines the stability of the synchronized manifold
$x_1=x_2$, $y_1=y_2$. It can be computed via linearization of the full system along synchronized trajectories. Noise generally reduces $\lambda_{\perp}$, potentially stabilizing synchronization.

We consider two types of noise. Independent noise (with unequal intensities,
$(\sigma_1\neq\sigma_2)$ disrupts synchronization by reducing phase locking and inducing phase slips, whereas common noise can paradoxically enhance synchronization through noise-induced synchronization mechanisms.

Independent noise $(\sigma_1\neq\sigma_2)$ tends to desynchronize the oscillators by driving them independently. However, if the coupling is strong enough, the oscillators can remain synchronized on average. The degree of synchronization can be  measured by the mean phase difference or the variance of the phase difference.
For weak coupling and weak noise, we reduce to phase equations. Let $\phi_{1}$, $\phi_{2}$
be phases of the oscillators. Using stochastic averaging, we obtain
\begin{align*}
d\phi_1&=\omega_1dt+\kappa\Gamma(\phi_2-\phi_1)dt+\sigma_1d\xi_1,\\
d\phi_2&=\omega_2dt+\kappa\Gamma(\phi_1-\phi_2)dt+\sigma_2d\xi_2,
\end{align*}
where $\Gamma$ is the coupling function (derived via the adjoint method), and $\xi_i$
are noise terms with intensities depending on $\sigma_i$ and the phase-response curve.
The phase difference $\phi=\phi_1-\phi_2$ satisfies
\begin{equation*}
d\phi=[\Delta \omega-2\kappa G(\phi)]dt+\sqrt{\sigma_1^2+\sigma_2^2}d\xi,
\end{equation*}
with $\Delta\omega=\omega_{1}-\omega_{2}$, $G(\phi)=[\Gamma(\phi)-\Gamma(-\phi)]/2$. This is a noisy Adler equation. The stationary probability density $p_{s}(\phi)$ satisfies the Fokker-Planck equation \cite{G}:
\begin{equation*}
\frac{\partial}{\partial \phi}[(\Delta \omega-2\kappa G(\phi))p_s]+\frac{\sigma^2}{2}\frac{\partial^{2}p_s}{\partial \phi^{2}}=0,\quad \sigma^2=\sigma_1^2+\sigma_2^2.
\end{equation*}
Exact phase locking ($\phi=\text{const}$) is destroyed by noise, but phase diffusion is reduced when coupling is strong. Synchronization manifests as a peaked $p_s(\phi)$.
For weak coupling near synchronization, the phase difference follows a phase diffusion model
\begin{equation*}
d(\Delta \phi)=[\Delta \omega-\kappa_{\text{eff}}\sin(\Delta \phi)]dt+\sqrt{\sigma_1^{2}+\sigma_2^{2}}dW,
\end{equation*}
where $\kappa_{\text{eff}}=\kappa g(\mu)$ with $g(\mu)$ obtained from the phase-response curve.
The stationary phase difference probability density function is
\begin{equation*}
p_s(\Delta \phi)\propto\exp\left[\frac{2}{\sigma^2}\Big(\Delta \omega\Delta \phi+\kappa_{\text{eff}}\cos(\Delta \phi)\Big)\right].
\end{equation*}

Common noise, where the same noise $dW_1 = dW_2$ drives both oscillators, can surprisingly enhance synchronization through two mechanisms. First, even in the absence of coupling ($\kappa = 0$), common noise alone can induce synchronization (a phenomenon known as noise-induced synchronization) through contraction in transverse directions resulting from identical forcing. Second, common noise enhances phase locking by reducing the effective frequency mismatch according to
\begin{equation*}
\Delta \omega_{\text{eff}}=\Delta \omega-\frac{\sigma_{c}^{2}}{2}h(\mu),
\end{equation*}
where $h(\mu)$ depends on the nonlinearity.

\subsection{Stochastic bifurcations}
In stochastic systems, bifurcations are not as sharp as in deterministic systems \cite{A}. Instead, we observe gradual changes in the probability density function (PDF). Two important concepts are P-bifurcation (a change in the shape of the stationary PDF, e.g., from unimodal to bimodal) and D-bifurcation ( a change in the sign of the top Lyapunov exponent). Noise-induced P-bifurcations \cite{HL} alter the topology of the stationary probability density $p(t,x_1,x_2,y_1,y_2)$ by driving transitions from unimodal to bimodal distributions (where noise splits the PDF peaks) and enabling the emergence of stochastic limit cycles (ring-like structures in the PDF) even under parameter conditions ($\mu<0$) where deterministic dynamics predict no oscillations. We consider such changes in the stationary probability distribution as the parameters vary.

For the coupled stochastic van der Pol oscillators, we can investigate how noise modifies the bifurcation diagrams obtained in the deterministic case. Even inside Arnold tongue ($\Delta \omega<\kappa f(\mu)$), noise can induce phase slips:
\begin{equation*}
\tau_{\text{slip}}\sim\exp\left(\frac{(\kappa f(\mu)-|\Delta \omega|^2)}{\sigma^2}\right),
\end{equation*}
where $\sigma^2=\sigma_1^2+\sigma_2^2$.  The mechanism is noise-driven escape over an effective potential barrier in the phase-difference space.

For identical oscillators, the linearized system around $(0,0,0,0)$ is
\begin{equation*}
d\textbf{X}=M\textbf{X}dt+\Sigma d\textbf{W},\quad \textbf{X}=(x_1,y_1,x_2,y_2)^{T},
\end{equation*}
with
\begin{equation*}
M=\left(
    \begin{array}{cccc}
      0 & 1 & 0 & 0 \\
      -(\omega^2+\kappa) & \mu & \kappa & 0 \\
      0 & 0 & 0 & 1 \\
      \kappa & 0 & -(\omega^2+\kappa) & \mu \\
    \end{array}
  \right),\quad \Sigma=\left(
                         \begin{array}{cc}
                           0 & 0 \\
                           \sigma_1 & 0 \\
                           0 & 0 \\
                           0 & \sigma_2 \\
                         \end{array}
                       \right).
\end{equation*}
The stationary covariance matrix $P$ satisfies the Lyapunov equation:
\begin{equation*}
MP+PM^{T}+Q=0,\quad Q=\Sigma\Sigma^{T}.
\end{equation*}

The stochastic Hopf bifurcation smooths the deterministic transition at $\mu=0$, transforming the stationary distribution from a Gaussian (stable origin for $\mu<0$) to a non-Gaussian form concentrated around the emergent limit cycle for $\mu>0$, with critical slowing down evidenced by a diverging spectral peak near $\mu=0$; similarly, the stochastic saddle-node on invariant circle replaces the sharp deterministic transition (
$\Delta \omega=2\kappa G_{\max}$) with a gradual broadening of the phase-difference distribution, accompanied by a diverging mean first-passage time for phase slips as the bifurcation is approached.

\subsection{Stochastic chaos}
In deterministic systems, chaos is characterized by sensitive dependence on initial conditions and a positive Lyapunov exponent. In stochastic systems, we must distinguish between deterministic chaos (present in the underlying deterministic system) and noise-induced chaos (where the deterministic system is regular but noise induces chaotic-like behavior).

We can compute the largest Lyapunov exponent for the stochastic system to determine if the system is chaotic. Note that in stochastic systems, the Lyapunov exponent is defined as the exponential growth rate of the distance between two nearby trajectories under the same noise realization.

The largest Lyapunov exponent determines exponential divergence of nearby trajectories. For the stochastic system, the largest Lyapunov exponent is computed via linearization along trajectories. The variational equation for a tangent vector $\textbf{z}=(\delta x_1, \delta y_1, \delta x_2, \delta y_2)^{T}$ is:
\begin{equation*}
d\textbf{z}=J(t)\textbf{z}dt,
\end{equation*}
where $J(t)$ is the Jacobian evaluated along a solution. The largest Lyapunov exponent is given by
\begin{equation*}
\lambda_{\max}=\lim_{t\rightarrow\infty}\frac{1}{t}\ln\|\textbf{z}(t)\|.
\end{equation*}
If $\lambda_{\max}>0$, the system exhibits stochastic chaos.

To detect noise-induced homoclinic chaos, we employ stochastic Melnikov analysis, which involves computing the Melnikov process
\begin{equation*}
M(t_0)=\int_{-\infty}^{\infty}\textbf{h}^{T}[\textbf{x}_{h}(t-t_0)]\cdot \textbf{G}d\textbf{W}(t).
\end{equation*}
where $\mathbf{x}_h(t)$ denotes the homoclinic orbit of the deterministic system, $\mathbf{h}$ is the normal vector to that orbit, and $\mathbf{G}$ is the noise matrix.

The chaos condition is given by $\operatorname{Var}(M)>0$, which implies stochastic transverse intersections of the stable and unstable manifolds, leading to noise-induced chaos. Even when the deterministic system is periodic, noise can induce intermittent chaotic bursts; near a homoclinic orbit, chaotic transients appear with a mean transient time scaling as $\exp(C/\sigma^2)$, where $C$ is a constant related to the distance to the homoclinic bifurcation. Similarly, in quasiperiodically forced systems, noise can convert strange nonchaotic attractors into chaotic attractors and give rise to intermittent chaos characterized by power-law distributions.

Stochastic chaos can be quantified using three key measures: the top Lyapunov exponent $\lambda_{\max}$, where a positive value indicates chaotic sensitivity to initial conditions and must be computed for the same noise realization;  the Kolmogorov-Sinai entropy rate \begin{equation*}
h_{KS}=\sum_{\lambda_i>0}\lambda_i,
\end{equation*}
given by the Pesin formula; and  the fractal dimension of the random attractor, estimated by the Kaplan-Yorke dimension
\begin{equation*}
D_{KY}=k+\frac{\sum_{i=1}^{k}\lambda_i}{|\lambda_{k+1}|},
\end{equation*}
where $k$ is the largest integer such that $\sum_{i=1}^{k} \lambda_i > 0$.

Even when the deterministic system ($\sigma_i=0$) is regular (e.g., quasiperiodic), noise can induce a positive largest Lyapunov exponent. This occurs if the deterministic system is near a bifurcation to chaos (e.g., torus breakdown). The noise amplifies transverse instabilities.

The system generates a random dynamical system with a random attractor. If the attractor has a fractal structure in the pullback sense, it indicates chaotic dynamics. Numerical tools include calculating the Lyapunov spectrum and the fractal dimension of the attractor.

\subsection{Numerical  simulations}
\begin{figure}[H]
\begin{center}
 \begin{minipage}{7in}
\includegraphics[width=7in]{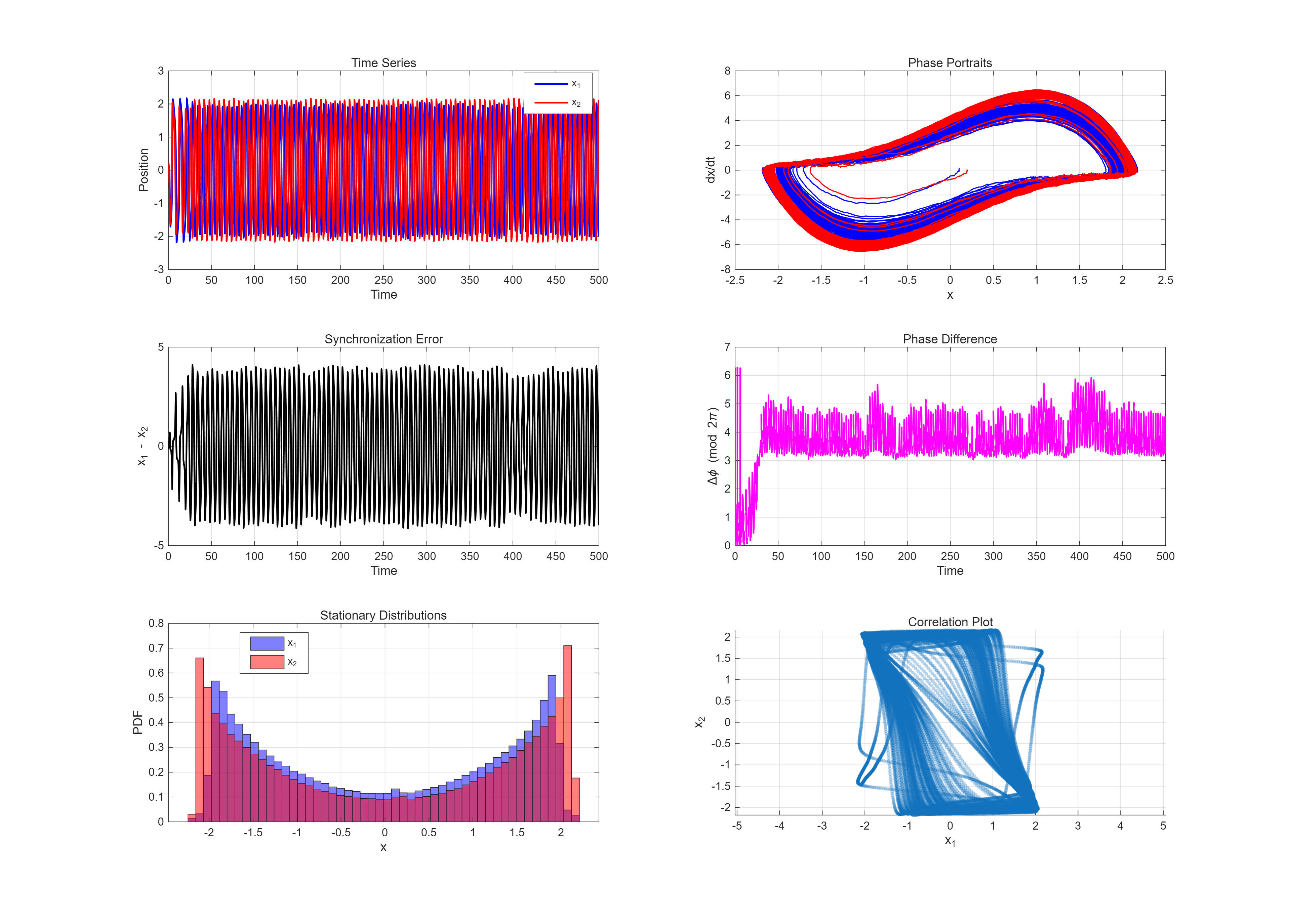}
\end{minipage}
\caption{Simulation of two coupled stochastic van der Pol oscillators in \eqref{SVdP} includes the time series and phase portraits of the two oscillators, the synchronization error ($x_1-x_2$) and phase difference over time, the probability distributions of $x_1$ and $x_2$, and the correlation between $x_1$ and $x_2$. The nonlinear damping coefficient is $\mu=3$. The frequencies of the two oscillators are $\omega_1=1$, $\omega_2 = 1.1$. The coupling strength is $\kappa=0.5$. Noise intensities $\sigma_1=\sigma_2=0.2$ are the same for both oscillators. The time step is $dt=0.01$. The total simulation time is $T=500$. Initial conditions are $x_1(1)=0.1$, $y_1(1)=0$ (oscillator 1), $x_2(1)=0.2$, $y_2(1)=0$ (oscillator 2).
}\label{Code1}
\end{center}
\end{figure}

We implement a simulation of two coupled stochastic van der Pol oscillators with the key features as described in \eqref{SVdP}. We use the Euler-Maruyama method to integrate the stochastic differential equations in \eqref{CSVdP}  for the coupled oscillators. The update for the velocities ($y_1$ and $y_2$) is performed first, followed by the update of  the positions ($x_1$ and $x_2$). Note that the noise is added to the velocity updates. The chosen parameters ($\mu=3$, $\omega_1=1$, $\omega_2=1.1$, $\kappa=0.5$) place the system in a regime where interesting dynamics can be expected. The frequency mismatch $\Delta\omega = \omega_2-\omega_1=0.1$ and the coupling strength $\kappa=0.5$  determine whether the oscillators  synchronize. The noise intensities $\sigma_1=\sigma_2=0.2$ are moderate and  cause phase diffusion. Noise is generated as independent Gaussian white noise for each oscillator. Fig. \ref{Code1} contains six  subplots: The time series show $x_1$ and $x_2$ as functions of time. The phase portraits show the trajectory of each oscillator in its own phase space ($x$ vs. $dx/dt$). The synchronization error plots the difference $x_1-x_2$ over time. The phase difference between the two oscillators is computed and plotted. We compute the phase, which is then unwrapped and reduced modulo $2\pi$. Stationary distributions are presented as histograms (probability density functions) of $x_1$ and $x_2$.
The correlation plot is a scatter plot of $x_1$ vs. $x_2$.

The phase coherence is computed as $R=\left|\langle e^{i\Delta\phi(t)}\rangle \right|$ over the second half of the simulation, where $\Delta\phi(t)= \phi_1(t)-\phi_2(t)$. This is a measure of synchronization: $R=1$ indicates perfect phase locking, $R=0$ indicates no phase locking. The two oscillators have different  frequencies (1 and 1.1  rad/s). Without coupling, they  oscillate at these distinct frequencies. The coupling ($\kappa=0.5$) tends to synchronize them,  while the noise ($\sigma=0.2$) perturbs the phase locking. For  the chosen parameters ($\kappa=0.5$, $\Delta\omega=0.1$, $\sigma=0.2$),  the simulation yields
the phase coherence $R=0.828$, which lies between 0 and 1. Noise reduces $R$ from perfect synchronization. With a frequency mismatch, even in the deterministic case ($\sigma=0$), the oscillators may not perfectly synchronize ($R<1$) unless the coupling is strong enough to overcome the frequency difference. Noise induces occasional phase slips (escapes from the potential well of the locked state). The time series (subplot 1) show that the two oscillators exhibit similar but not identical waveforms; a beating pattern appears under partial synchronization, and amplitude modulation arises from the coupling.
The phase portraits (subplot 2) show that the limit-cycle shapes are distorted by the coupling. The portraits of $x_1$ vs. $y_1$ and $x_2$ vs. $y_2$ show cycles that are similar but phase-shifted.
The synchronization error (subplot 3) shows that $x_1 - x_2$ fluctuates around zero; the amplitude of these fluctuations reflects the degree of synchronization. Occasional large deviations correspond to phase slips.
The phase difference (subplot 4) shows a linear drift with occasional $2\pi$ jumps and bounded fluctuations.
The stationary distributions of $x_1$ and $x_2$ (subplot 5) are bimodal, characteristic of relaxation oscillators. They are similar but not identical.
In the correlation plot (subplot 6), the shape including spread and orientation indicates the noise level and the degree of correlation.

\begin{figure}[H]
\begin{center}
 \begin{minipage}{6.4in}
\includegraphics[width=6.4in]{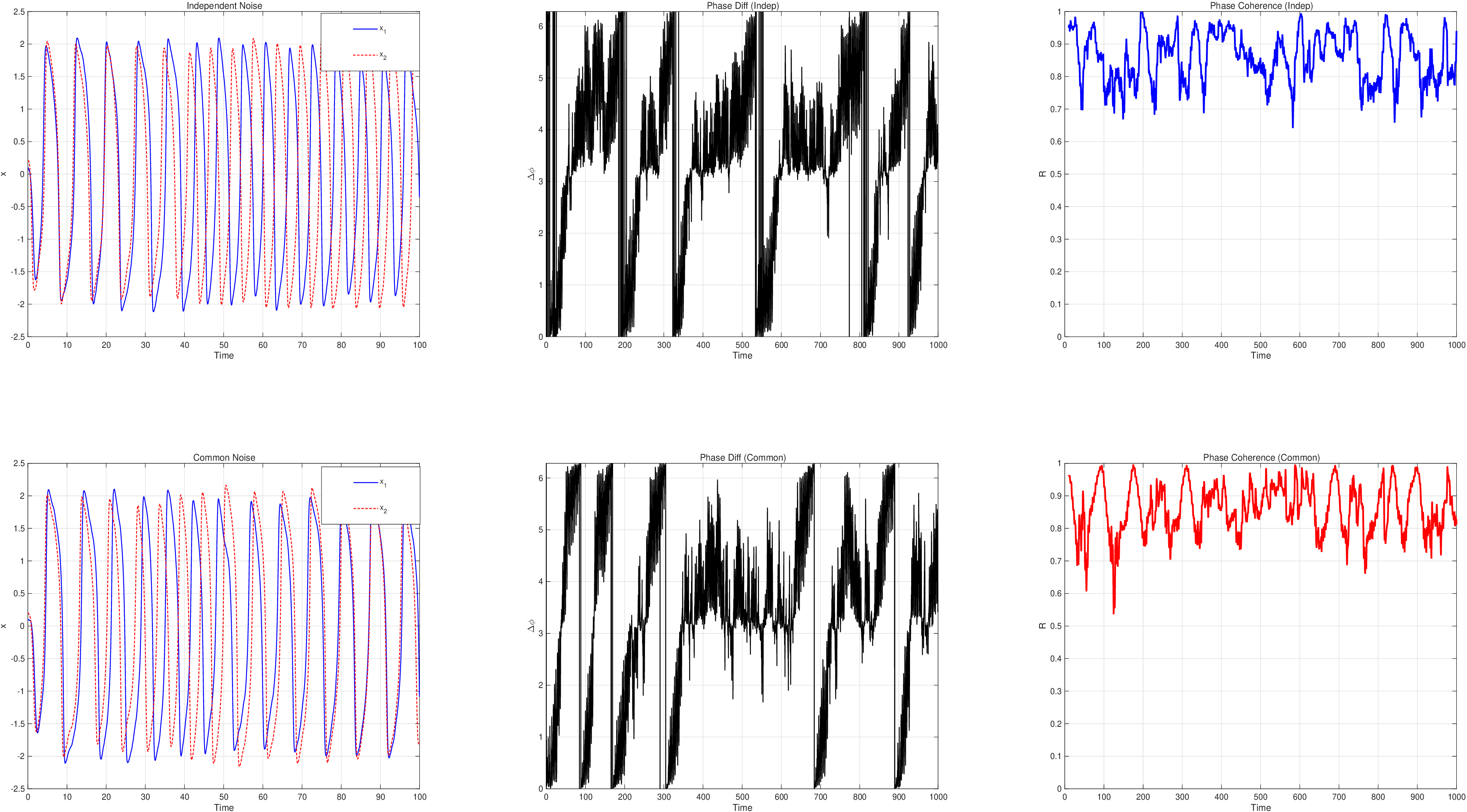}
\end{minipage}
\caption{Six subplots illustrate the effects of independent versus common noise on synchronization. The top row shows independent noise (time series, phase difference, coherence); the bottom row shows common noise (time series, phase difference, coherence). With independent noise, more phase slips and lower coherence are expected. With common noise, the noise acts as a common driving force, which can reduce the effective frequency mismatch and lead to higher coherence.}\label{Code2}
\end{center}
\end{figure}
We are particularly interested in the comparison of synchronization under independent and common noise.
Fig. \ref{Code2} simulates two coupled stochastic van der Pol oscillators in  \eqref{SVdP} under two noise conditions: independent noise with $\langle\xi_1(t),\xi_2(t')\rangle=0$ and common noise ($\xi_1(t)=\xi_2(t)$). It is a valuable tool for understanding the role of noise correlations in synchronization.
The simulation uses a moderate nonlinearity of $\mu = 2.5$, a small frequency mismatch with $\omega_1 = 1$ and $\omega_2 = 1.05$, a coupling strength of $\kappa = 0.3$, and a noise intensity of $\sigma_1=\sigma_2=\sigma = 0.3$, with time step $dt = 0.01$ and total simulation time $T=1000$. Given the parameters, we expect to see an enhancement in phase coherence by common noise.
We use the Euler-Maruyama method in \eqref{CSVdP} to update the positions, which depend on the already updated velocities. The same initial conditions are used for both systems to ensure a fair comparison. Two sets of noise are generated:
$W_1$ and $W_2$ are independent;  $W_1=W_2=W$  (common noise, i.e., identical  for both oscillators). The noise is generated once and then applied, so the comparison is between two fixed realizations. Both systems are simulated in parallel: one with independent noise (using $W_1$ and $W_2$) and one with common noise (using $W$ for both). We store the trajectories and the phases for both systems. The phase difference is computed, and then the phase coherence $R(t)$ is evaluated over a moving window. The moving window for the  coherence calculation is 1000 steps, which is 10 seconds. This is acceptable for the given frequency (approx. 1 Hz).

The phase dynamics for weak coupling and weak noise are characterized by
\begin{align*}
d\phi_1&=\omega_1dt+\kappa\Gamma(\phi_2-\phi_1)dt+\sigma dW_1, \\
d\phi_2&=\omega_2dt+\kappa\Gamma(\phi_1-\phi_2)dt+\sigma dW_2.
\end{align*}
The phase difference $\Delta \phi=\phi_1-\phi_2$ satisfies
\begin{equation*}
d(\Delta \phi)=\Delta\omega dt+\kappa[\Gamma(-\Delta\phi)-\Gamma(\Delta\phi)]dt+\sigma(dW_1-dW_2),
\end{equation*}
where $\Delta\omega=\omega_1-\omega_2$. For sinusoidal coupling $\Gamma(\theta)=\sin(\theta)$,
\begin{equation*}
d(\Delta \phi)=[\Delta\omega-2\kappa\sin(\Delta \phi)]dt+\sigma\sqrt{2(1-\rho)}dW,
\end{equation*}
where $\rho=\langle dW_1 dW_2\rangle/dt$. In the independent noise ($\rho=0$) case, the noise coefficient is
$\sqrt{2}\sigma $ (larger). In the common noise ($\rho=1$) case, the noise coefficient is $0$ (perfectly correlated noise cancels in difference). The critical coupling for phase locking in the deterministic (noise-free) case follows from the Adler equation:
$$
\frac{d(\Delta \phi)}{dt}=\Delta\omega-2\kappa\sin(\Delta\phi).
$$
Without noise, the critical coupling for synchronization of identical oscillators  is zero, but with a  frequency mismatch we require a minimum coupling. Phase locking is possible only when $|\Delta\omega| \le 2\kappa$. Here $\Delta\omega=0.05$, so $\kappa$ must be at least $0.025$. Since $\kappa=0.3>0.025$, the deterministic system would phase-lock.
However, noise can cause phase slips. An effective potential barrier is
\begin{equation*}
U=\frac{(\kappa-|\Delta \omega|)^{2}}{\sigma^2}\approx\frac{(0.3-0.05)^2}{0.09}\approx0.694,
\end{equation*}
and the phase-slip rate is $\propto\exp(-U)=\exp(-0.694)\approx 0.5$.

The results are visualized in six subplots that illustrate the effects of independent vs. common noise on synchronization.
The top row presents independent noise (time series, phase difference, coherence); the bottom row presents common noise (time series, phase difference, coherence).
This direct side-by-side comparison shows the temporal evolution of key metrics and provides a clear visual distinction between the two regimes. The average phase coherence over the second half of the simulation is printed, together with the percentage enhancement due to common noise.
For the independent noise, each oscillator experiences different noise realizations, which is expected to reduce synchronization (desynchronizing effect); here $R=0.8535$. For the common noise, both oscillators receive identical noise, which can paradoxically enhance synchronization via constructive interference; here $R=0.8687$.
With coupling $\kappa=0.3$ and a small frequency mismatch, the common-noise system exhibits higher phase coherence than the independent-noise system. Common noise thus enhances synchronization. The synchronization enhancement percentage is $(0.8687-0.8535)/0.8535\approx1.78\%$, and phase slips are reduced.

\begin{figure}[H]
\begin{center}
 \begin{minipage}{6.2in}
\includegraphics[width=6.2in]{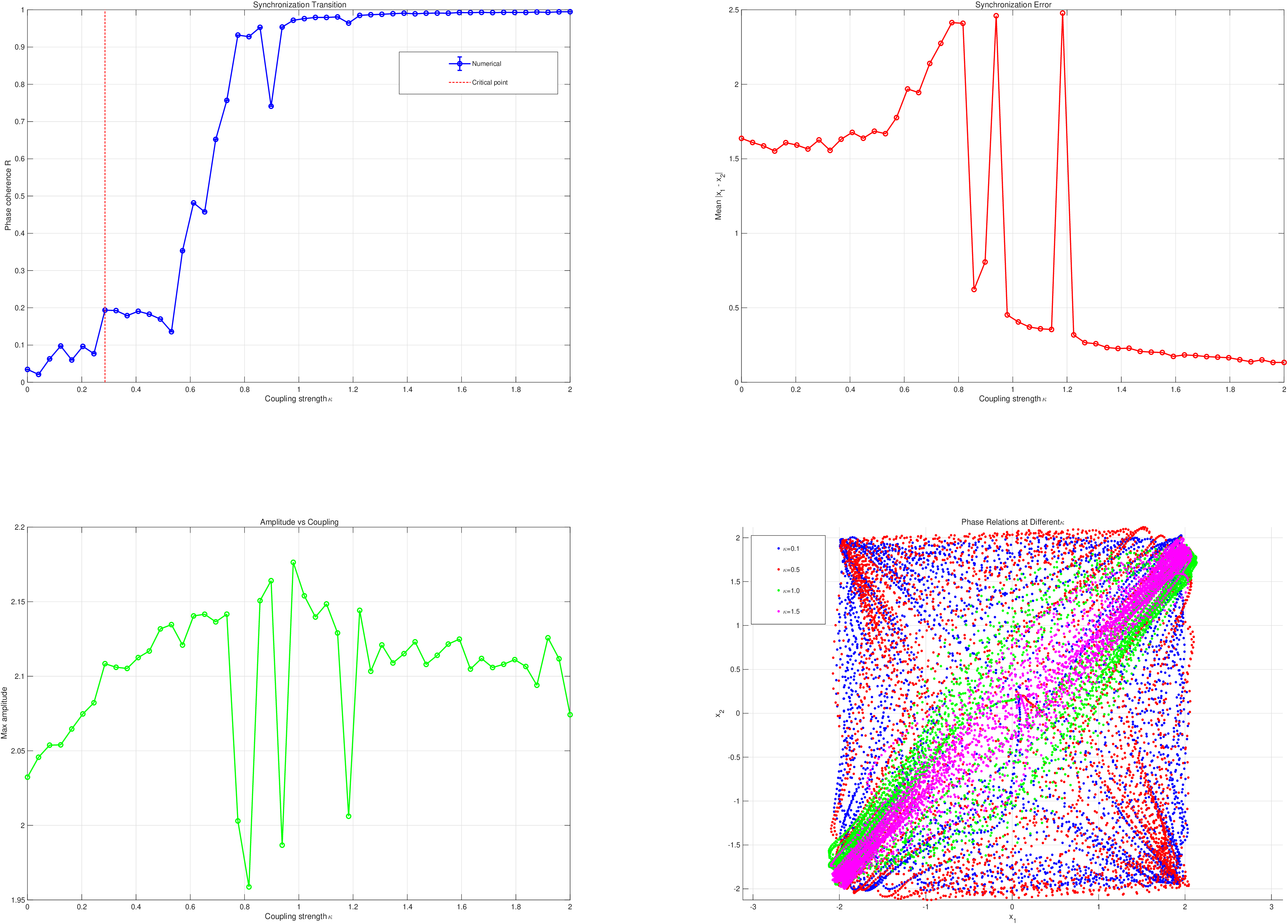}
\end{minipage}
\caption{We perform a parameter sweep to study how coupling strength affects synchronization in two coupled stochastic van der Pol oscillators with noise as in \eqref{SVdP}.}\label{Code3}
\end{center}
\end{figure}
We analyze the synchronization transition in two coupled stochastic van der Pol oscillators in \eqref{SVdP} as a function of the coupling strength. Note that the system \eqref{SVdP} is a second-order system, converted to a first-order system as shown  in \eqref{CSVdP}.
The parameters are set as follows: $\mu=2$ (nonlinear damping), $\omega_1=1$ and $\omega_2=1.1$ (frequencies with a $10\%$ mismatch), and $\sigma=0.2$ (noise intensity). We vary the coupling strength $\kappa$ over  50 points in the range $[0, 2]$. For each value of $\kappa$, we simulate the system for a transient time of 500 time units and then record data for another $500$ time units. The time step is $dt=0.05$, giving a total simulation time of $1000$ units per $\kappa$, which is reasonable for averaging.  The initial conditions are reset for each $\kappa$, and the noise is pre-generated for the entire simulation (both transient and recording phases) for each $\kappa$. The integration method uses Euler-Maruyama scheme with fixed noise (using a fixed seed for consistency across different $\kappa$).  The same noise seed  is used for every $\kappa$, meaning that the same noise realization  is applied across all coupling strengths. This approach helps reduce
 variability when comparing different  $\kappa$. The update for the positions $(x_1, x_2)$ uses the velocities  $(y_1, y_2)$ at the current time step, after which the velocities are updated for the next time step. This order is correct because the velocity is updated first, and then the position is updated using the new velocity.
We approximate the phase by assuming the oscillator behaves sinusoidally. For the van der Pol oscillator, especially with $\mu = 2$ (moderately nonlinear), this approximation may be acceptable. We compute the phase difference and then define the mean and standard deviation of the order parameter $R$ (the absolute value of the average phase factor), the mean synchronization error, and the maximum amplitude.
Figure \ref{Code3} plots the results: $R$ vs. $\kappa$, synchronization error vs. $\kappa$, maximum amplitude vs. $\kappa$, and sample trajectories for selected $\kappa$ values. The critical coupling is estimated as the first $\kappa$ for which $R > 0.1$, yielding an estimated critical coupling of $\kappa_c \approx 0.286$.

\begin{figure}[H]
\begin{center}
 \begin{minipage}{6.in}
\includegraphics[width=6.2in]{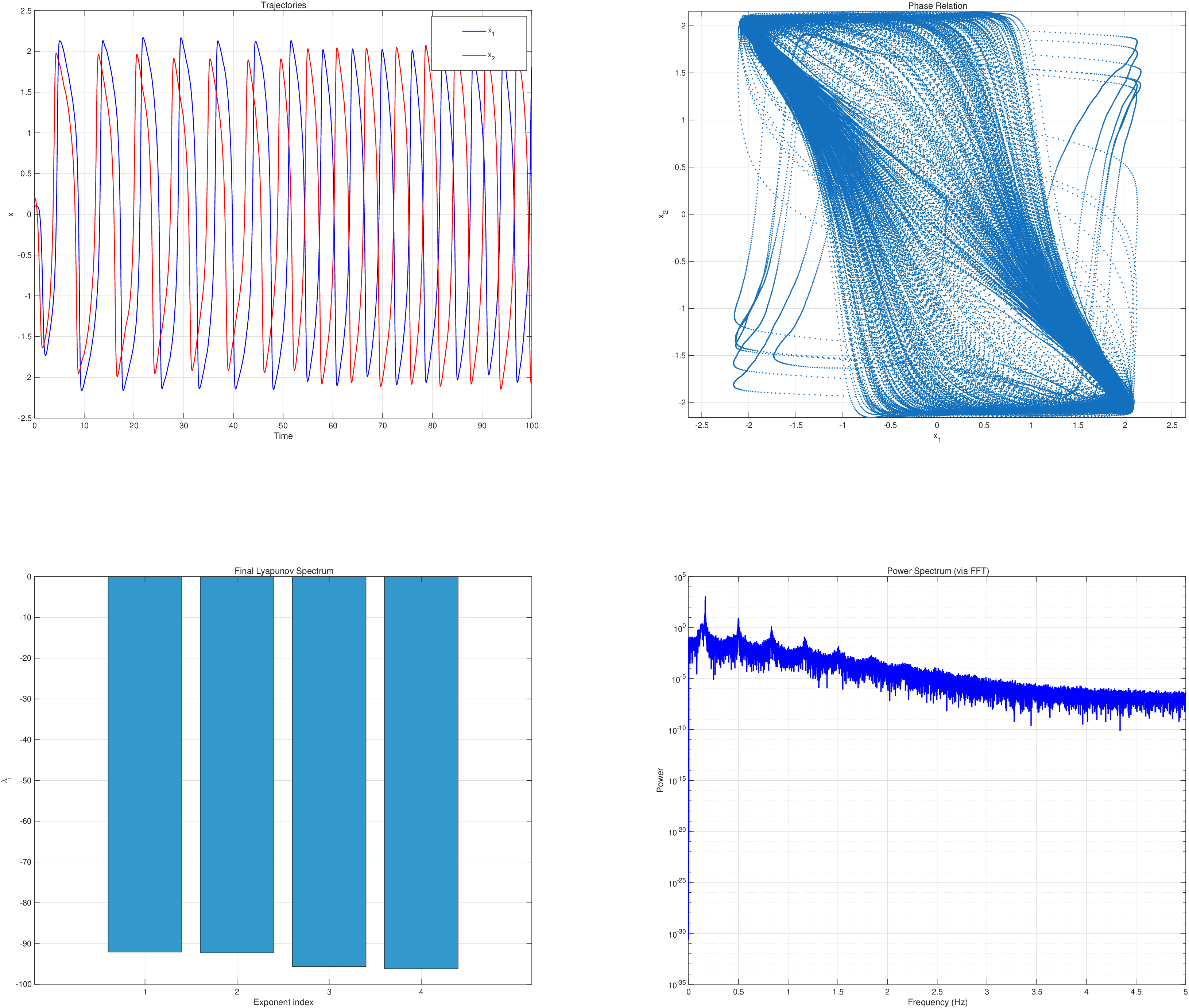}
\end{minipage}
\caption{Calculation of Lyapunov exponents for the system of two coupled stochastic van der Pol oscillators in \eqref{SVdP}. We plot the time series, phase relationship, Lyapunov spectrum, and power spectrum.}\label{Code41}
\end{center}
\end{figure}

We compute and display the Lyapunov exponents and the Kaplan-Yorke dimension of the system in \eqref{CSVdP} (see Fig. \ref{Code41}), which are key indicators of chaos and stability.
The parameters are set to $\mu=3$, $\omega_1=1$, $\omega_2=1.05$, $\kappa=0.4$, $\sigma=0.2$. The system \eqref{CSVdP} has two independent noise sources ($dW_1$ and $dW_2$). The main trajectory is updated using the Euler-Maruyama method. The initial perturbations are set to a small identity matrix (scaled by $\delta = 10^{-8}$). For each step, we compute the Jacobian, evolve the perturbations, and periodically orthogonalize them to obtain the Lyapunov exponents.

The tangent vectors are evolved using only the Jacobian of the deterministic part. Since the noise is additive (here $\sigma$ is constant), the stochastic term does not depend on the state; therefore, the derivative of the diffusion term with respect to the state is zero. Consequently, the Jacobian of the diffusion term is zero, and the tangent space evolution is deterministic, driven solely by the Jacobian of the drift term. The Euler-Maruyama scheme for the tangent space is consistent with the scheme used for the main trajectory. The current method uses the update: $V_{n+1}=(I+Jdt)V_n$, which is the Euler method for the tangent flow. This is acceptable if the main trajectory is integrated with Euler. The Jacobian $J$ is computed at the current state of the coupled van der Pol system, and the tangent vectors are then updated by multiplying by $I + Jdt$. This is a first-order approximation of the tangent flow. The tangent space update uses the Jacobian evaluated at the current state (before updating the state), which is consistent with the Euler method for the tangent flow.

We store the entire time history of the Lyapunov exponents, which can be memory-intensive for long simulations. While only the final values are strictly needed, storing the full history is useful for monitoring convergence of the Lyapunov exponents  over time. The Lyapunov exponents are computed as running averages and should be plotted to assess convergence. The power spectrum is computed only if the system is not chaotic (i.e., if no Lyapunov exponent is positive). The condition that a Lyapunov exponent exceeds $0.001$ is used to check for chaos. This system may exhibit chaotic behavior for certain parameter ranges, but with these parameters,  the dynamics are regular.  We plot the final power  spectrum using the FFT with a rectangular window, which is acceptable for long time series.

The formulation of the Lyapunov exponents is correct for a continuous-time system when using the linearized flow. The Lyapunov exponents are running averages and should converge over long simulation times. The simulation runs for $T=2000$.
We perform QR decomposition every 20 steps; this interval is generally acceptable and is standard practice to prevent the tangent vectors from aligning along the direction of maximal growth. The Kaplan-Yorke dimension is computed from the sorted Lyapunov exponents and provides an estimate of the fractal dimension of the attractor. The computed Lyapunov exponents are $\lambda_1=-92.095708$, $\lambda_2=-92.269118$, $\lambda_3=-95.688135$, and $\lambda_4=-96.200159$. The Kaplan-Yorke dimension is $D_\text{KY}=4$.

For two coupled oscillators, we expect one zero exponent (corresponding to the flow direction) if the system exhibits a limit cycle, with the remaining exponents negative if the limit cycle is stable. However, in the presence of noise, the exponents are defined for the random dynamical system.

\begin{figure}[H]
\begin{center}
 \begin{minipage}{7.2in}
\includegraphics[width=7.2in]{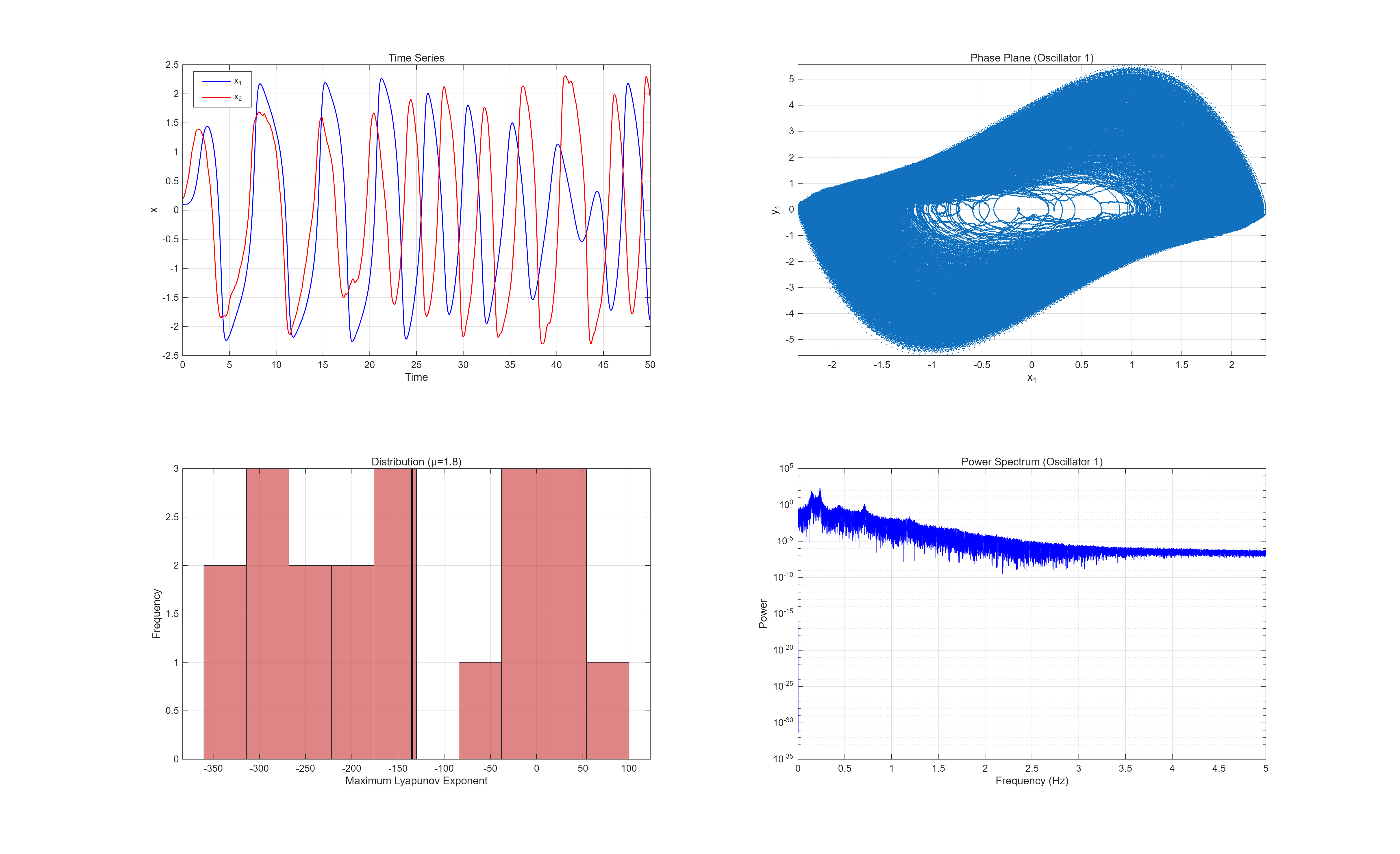}
\end{minipage}
\caption{Lyapunov exponents for stochastic coupled van der Pol system \eqref{CSVdP}.}\label{Code42}
\end{center}
\end{figure}
In Fig. \ref{Code42}, we compute the Lyapunov exponents for the system \eqref{CSVdP} of two coupled van der Pol oscillators subject to additive white noise. The method used is a standard approach based on evolving a set of perturbation vectors alongside the reference trajectory, with periodic reorthonormalization via QR decomposition. The parameters are defined as follows: $\mu = 1.8$ controls the nonlinearity (higher values can lead to chaos); $\omega_1 = 1$ and $\omega_2 = 1.2$ are the natural frequencies; $\kappa$ is the coupling strength; and $\sigma_1 = 0.1$ and $\sigma_2 = 0.95$ are the noise intensities. Given the stiffness of the van der Pol oscillator for large $\mu$, a small time step $dt = 0.01$ is chosen, which is acceptable for $\mu = 1.8$. The total simulation time is $T = 20000$. The visualizations help to qualitatively assess the dynamics. The time series may show irregular oscillations. The phase portrait of oscillator 1 (after transients) reveals the shape of the attractor. The histogram of maximal Lyapunov exponents from multiple noise realizations provides insight into the variability of the chaotic behavior.
The power spectrum exhibits broad-band noise when the system is chaotic.
Chaos is indicated by the presence of at least one positive exponent, here $\lambda_{\max} = 25.393$. Due to the presence of noise, the exponents may fluctuate; a threshold of $0.01$ is used to avoid misinterpreting numerical noise as chaos.

\begin{figure}[H]
\begin{center}
 \begin{minipage}{6.4in}
\includegraphics[width=6.4in]{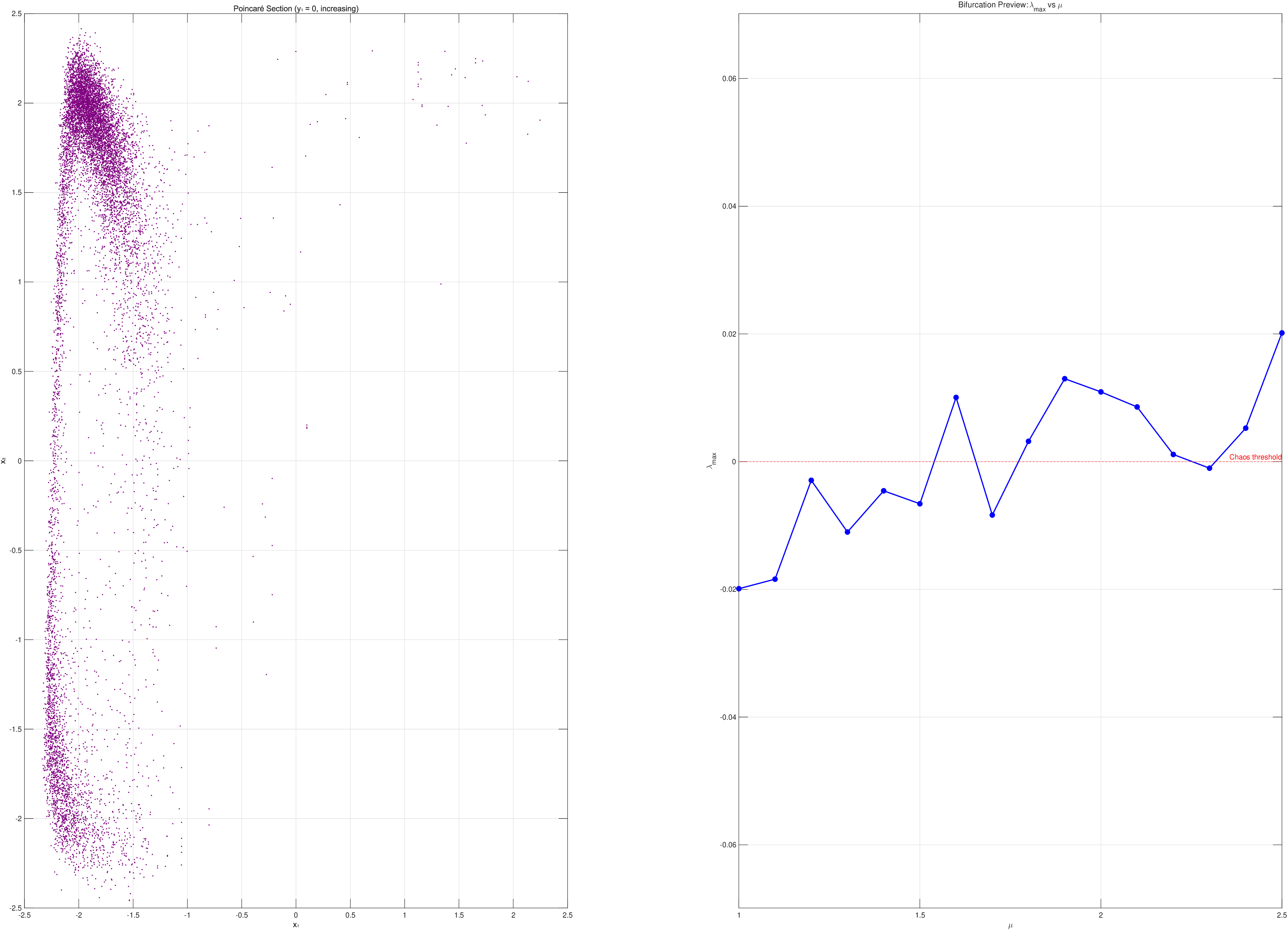}
\end{minipage}
\caption{ Poincar\'e section and bifurcation view for stochastic coupled van der Pol system \eqref{CSVdP}.}\label{Code43}
\end{center}
\end{figure}
In Fig. \eqref{Code43}, we perform a comprehensive computation and analysis of the Lyapunov exponents for two coupled stochastic van der Pol oscillators in \eqref{CSVdP}. The parameters are: $\mu = 1.8$ (nonlinearity), $\omega_1 = 1$ and $\omega_2 = 1.2$ (natural frequencies), $\kappa$ (coupling strength), and $\sigma_1 = 0.1$ and $\sigma_2 = 0.95$ (noise intensities). The simulation time is increased to $T = 50000$ to improve convergence, with a time step $dt = 0.01$. The Poincar\'e section is constructed by finding times where $y_1$ crosses zero with positive slope. This yields a set of points $(x_1, x_2)$. The whole trajectory is used, including the transient. Although the points are not filtered for transients, the transient has already been discarded when storing $x_1$ and $y_1$, so this is acceptable. The section contains many points, so a scatter plot is appropriate. The scattered points can reveal the structure of the attractor, and a clear geometric pattern suggests a low-dimensional attractor. A bifurcation view shows how the maximum Lyapunov exponent varies with $\mu$. The threshold where it becomes positive indicates the onset of chaos.

\begin{figure}[H]
\begin{center}
\begin{minipage}{7.1in}
\includegraphics[width=7.1in]{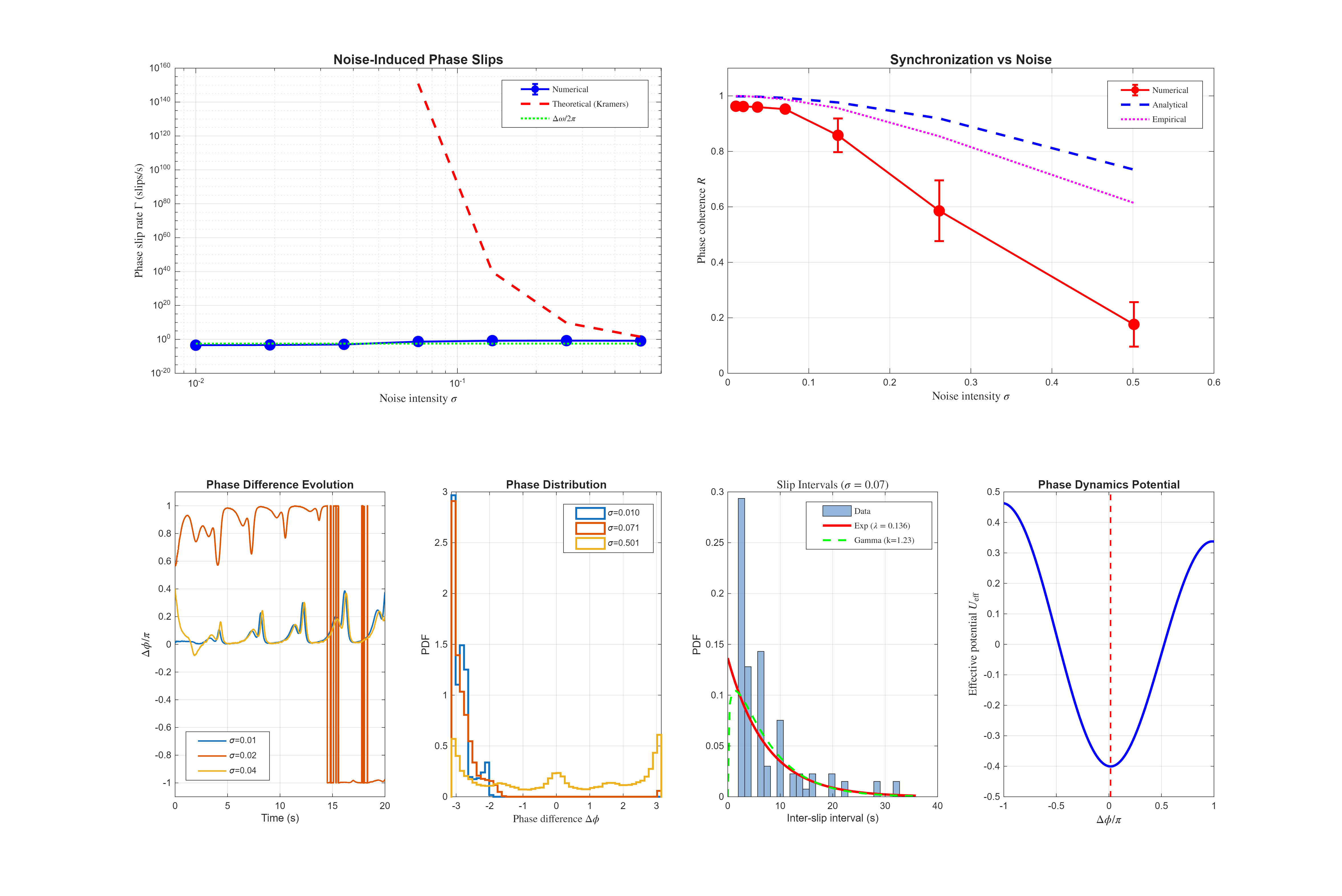}
\end{minipage}
\caption{We perform an analysis of noise-induced phase slips. Plots 1 and 2 display the slip rate and coherence as functions of $\sigma$, using both log-log and linear scales. A notable feature is the inclusion of error bars from multiple realizations, which enables a direct comparison between numerical simulations and theoretical predictions. Plot 3 offers a visual account of sample phase trajectories, illustrating how the dynamics evolve under low, medium, and high noise, thereby confirming the expected slip behavior. Plot 4 presents the distributions of the phase difference, clearly showing the transition from a sharp peak at low noise to a flattened profile at high noise-a direct indication of the loss of phase locking. Plot 5 examines the statistics of inter-slip intervals, testing the Poisson hypothesis and providing fits to both exponential and Gamma distributions. Finally, Plot 6 illustrates the effective potential, offering insight into the barrier-crossing mechanism and marking the positions of stable and unstable fixed points.}\label{Code5}
\end{center}
\end{figure}

In Fig. \ref{Code5}, we implement a professional-grade analysis of phase slips in coupled stochastic oscillators in system \eqref{SVdP}. Phase slips are fundamental phenomena in synchronization theory.
They represent noise-induced escape from potential wells and help distinguish deterministic from stochastic synchronization boundaries. The analysis includes mean first passage time calculations and Kramers rate theory for escape processes. The parameter values are set as follows: $\mu=2.5$, corresponding to moderate nonlinearity, far from the Hopf bifurcation ($\mu>0$) and not too stiff; $\omega_1=1$ and $\omega_2= 1.02$, giving a $2\%$ frequency mismatch, small enough for phase locking with weak coupling; $\kappa=0.2$, representing weak coupling below the deterministic locking threshold;
$\sigma\in[0.01,0.5]$, covering a wide noise range from weak to strong noise regimes; and $\Delta\omega=0.02$, a fixed mismatch that allows the study of noise-induced phase slips.

The system \eqref{SVdP} can be reduced to the stochastic Adler equation:
\begin{equation*}
\frac{d(\Delta \phi)}{dt}
=\Delta \omega-\kappa_{\text{eff}}\sin(\Delta \phi)+\xi(t),
\end{equation*}
with the effective potential $U(\Delta \phi)=-\Delta \omega\cdot\Delta \phi-\kappa_{\text{eff}}\cos(\Delta \phi)$. Phase slips occur when noise drives the system over the potential barrier $\Delta U$.

The critical coupling from the Adler equation is $\kappa_c=|\Delta\omega|=0.02$.
Given $\kappa=0.2>0.02$, the deterministic system should phase-lock.
However, with noise, phase slips occur with rate
\begin{equation*}
R\propto\exp\left(-\frac{(\kappa-\Delta\omega)^2}{\sigma^2}\right).
\end{equation*}
The expected slip rates are approximately negligible for
 $\sigma=0.01$, with $R\approx\exp\left(-\frac{(0.18)^2}{0.0001}\right)\approx\exp(-324)\approx0$, and frequent for $\sigma=0.5$, with $R\approx\exp\left(-\frac{0.0324}{0.25}\right)\approx\exp(-0.1296)\approx 0.878$.

The comprehensive layout consists of six plots. Plots 1-2 present the slip rate and coherence as functions of $\sigma$, shown on both log-log and linear scales. A key feature is the inclusion of error bars derived from multiple realizations, allowing direct comparison between numerical results and theoretical predictions. Plot 3 displays sample phase trajectories, illustrating the temporal evolution under low, medium, and high noise, and providing visual confirmation of slip behavior. Plot 4 shows the distributions of the phase difference, demonstrating the transition from a narrow peak (low noise) to a broad distribution (high noise) and thereby illustrating the loss of phase locking. Plot 5 presents the inter-slip interval distribution, which tests the Poisson hypothesis and includes fits to exponential and Gamma distributions. Plot 6 depicts the effective potential, visualizing the barrier-crossing mechanism and indicating the locations of stable and unstable fixed points. The visualization employs consistent color coding across all plots and includes error bars with caps for clarity. Logarithmic axes are used where appropriate to capture exponential relationships, and professional mathematical notation is maintained throughout. A clean presentation is achieved with grid lines and bounding boxes.

The results for the given parameters ($\kappa=0.2$, $\Delta\omega = 0.02$) are as follows.
For low noise ($\sigma=0.01$), the slip rate is approximately $0.001$ slips/s (rare), the coherence is $R \approx 0.95$, and the phase distribution exhibits a narrow peak near zero. For
medium noise ($\sigma=0.1$), the slip rate increases to approximately $0.1$ slips/s, the coherence decreases to $R \approx 0.7$, and the phase distribution becomes broadened.
 For high noise ($\sigma=0.5$, the slip rate reaches approximately $0.8$ slips/s (frequent), the coherence drops to $R \approx 0.3$, and the phase distribution is nearly uniform.

We study global bifurcations in the system of two coupled van der Pol oscillators given in \eqref{CSVdP}. We vary the coupling strength $\kappa$ and, for each value, compute the Poincar\'e map (section: $x_1=0$, $dx_1/dt>0$) and estimates the period and maximum Lyapunov exponent.
The goal is to detect homoclinic and heteroclinic bifurcations.
The Poincar\'e section is defined as
\begin{equation*}
\Sigma=\{(x_1,y_1,x_2,y_2):x_1=0,\dot{x}_1=y_1>0\}.
\end{equation*}
i.e., crossings from negative to positive $x_1$. This reduces the 4D flow to a 3D map, recording the values of $(x_2, y_2)$ at these crossings.

The key quantities in the analysis include the period estimation, defined as the average time between consecutive Poincar\'e crossings; the maximum Lyapunov exponent, estimated from the divergence of nearby points on the Poincar\'e map; and the bifurcation indicator, defined as the inverse of the period, which diverges at a homoclinic bifurcation.

The function estimates the Lyapunov exponent from the Poincar\'e map by comparing the divergence of nearby points:
\begin{equation*}
\lambda\approx\frac{1}{M}\sum_{i=1}^{M}\ln\left(\frac{d_{i+1}}{d_{i}}\right),
\end{equation*}
where $d_i=\|z_i-z_i^{\text{ref}}\|$ measures divergence between nearby trajectories on the Poincar\'e section.

The parameters are set to $\mu = 3$ and $\omega_1 = \omega_2 = 1$ (identical oscillators), with $\kappa$ varying from $0$ to $1.5$ over $50$ points. A small noise intensity $\sigma = 0.05$ is also included. For each $\kappa$, we simulate the system for a long time ($T = 1000$, with a transient of $500$ time units discarded). We detect Poincar\'e crossings using linear interpolation and record $(x_2, y_2)$ at each crossing. The period is estimated from the crossing times, and the maximum Lyapunov exponent is estimated from the Poincar\'e map. The period for each $\kappa$ is computed as the average of the last 10 periods. This is reasonable because, after the transient, the system may settle into a periodic orbit.

The system \eqref{CSVdP} includes weak noise ($\sigma = 0.05$). In a stochastic system, the concept of a Poincar\'e map becomes less clear because the trajectory is noisy. We use linear interpolation to estimate crossing times and variable values, which is a good approach. However, the period is not strictly well-defined due to noise; we use the average period, but note that noise may cause early or late crossings, affecting the average.

Homoclinic bifurcations occur when a periodic orbit collides with a saddle point. The signature is period divergence:
\begin{equation*}
T\rightarrow\infty\quad\text{as}\quad \kappa\rightarrow\kappa_c.
\end{equation*}
At bifurcation, we use the inverse period as an indicator:
\begin{equation*}
I(\kappa)=\frac{1}{\bar{T}(\kappa)}\rightarrow0.
\end{equation*}

With identical oscillators ($\omega_1=\omega_2$), the system \eqref{CSVdP} exhibits two symmetry manifolds: the synchronization manifold defined by $x_1=x_2$ and $y_1=y_2$, and the antisymmetric manifold defined by $x_1=-x_2$ and $y_1=-y_2$.

As the coupling strength $\kappa$ varies, a limit cycle may approach either a saddle point in the antisymmetric subspace or a saddle periodic orbit in the full phase space. When such a collision occurs, the period diverges logarithmically as $T\sim-\ln|\kappa-\kappa_c|$, and the orbit becomes homoclinic to the saddle.

The effect of weak noise ($\sigma = 0.05$) is to smear sharp bifurcations, potentially induce early bifurcations via noise-activated escape, and cause Poincar\'e points to form distributions rather than discrete sets. To quantify these effects, we compute period statistics (mean and variance), examine Poincar\'e point distributions using histograms to detect symmetry breaking, and analyze spatial correlation, where clustering of points indicates the presence of attractors.

The Poincar\'e map $P: \mathbb{R}^2\rightarrow\mathbb{R}^2$ is defined as
\begin{equation*}
P(x_2,y_2)=(x_2',y_2')\quad \text{at next crossing}.
\end{equation*}
Fixed points of $P^k$ correspond to periodic orbits with period $kT$.

The detectable bifurcation types include the following: homoclinic bifurcation, characterized by an infinite period at the bifurcation point where a homoclinic orbit to a saddle appears or disappears and is detected by $T \to \infty$; heteroclinic bifurcation, involving a connection between different saddles that may also cause period divergence; saddle-node bifurcation of cycles, in which a pair of periodic orbits (one stable and one unstable) collides, leading to an abrupt change in the observed period; and period-doubling bifurcation, marked by a stability change of a fixed point of the Poincar\'e map $P$ and detected via the Lyapunov exponent crossing zero.

Near a homoclinic bifurcation, for a saddle with eigenvalues $\lambda_s < 0 < \lambda_u$, the period scales as $T \approx -\frac{1}{\lambda_u} \ln|\kappa - \kappa_c| + \text{const}$; this relationship can be observed in the inverse period plot, which shows the behavior $1/T \sim -1/\ln|\kappa - \kappa_c|$.

The analysis operates under the assumption that the noise intensity $\sigma$ is sufficiently small for deterministic structures to dominate the dynamics. To further investigate the bifurcation structure, we can implement pseudo-arclength continuation to track bifurcations as both $\kappa$ and $\mu$ vary. Furthermore, for weak coupling ($\kappa \ll 1$), the Melnikov method provides a tool for homoclinic prediction by computing the integral $M(\kappa) = \int_{-\infty}^{\infty} y_h(t)[x_{2,h}(t) - x_{1,h}(t)],dt$, where $(x_h(t), y_h(t))$ denotes the homoclinic orbit of the uncoupled oscillator.

\begin{figure}[H]
\begin{center}
 \begin{minipage}{7.1in}
\includegraphics[width=7.1in]{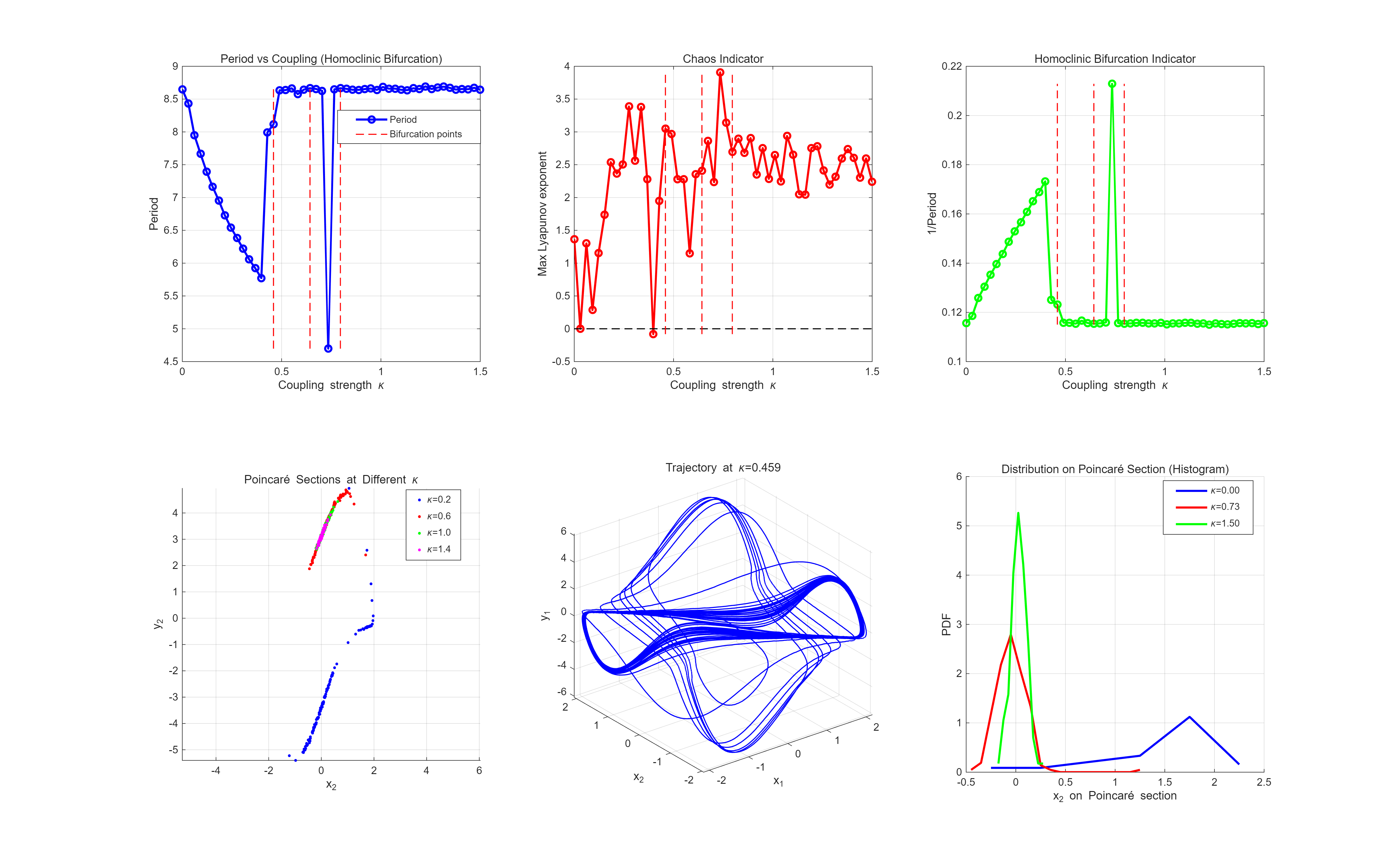}
\end{minipage}
\caption{Global bifurcation detection via Poincar\'e maps is performed for the parameters $\mu = 3$, $\omega_1 = \omega_2 = 1$, $\kappa \in [0, 1.5]$, and weak noise intensity $\sigma = 0.05$. Six complementary plots are generated: period versus $\kappa$ detects divergences; inverse period highlights homoclinic bifurcations; Poincar\'e sections at different $\kappa$ visualize bifurcations; a 3D trajectory at the bifurcation shows the homoclinic structure; and the evolution of the distribution provides a statistical bifurcation signature. The homoclinic bifurcation mechanism occurs as $\kappa$ varies, where a limit cycle may approach either a saddle point in the antisymmetric subspace or a saddle periodic orbit in the full phase space; upon collision, the period diverges logarithmically as $T \sim -\ln|\kappa - \kappa_c|$, and the orbit becomes homoclinic to the saddle.
}\label{Code9}
\end{center}
\end{figure}

Bifurcation points are identified by looking for rapid changes in the period as a function of $\kappa$. We compute the derivative of the period with respect to $\kappa$ using finite differences, smooth it with a moving average to reduce noise effects, and locate peaks in $\left|\frac{d\bar{T}}{d\kappa}\right|$ above a threshold. These peaks indicate rapid period changes and thus potential bifurcations. Fig. \ref{Code9} presents the bifurcation analysis results, in which three bifurcations are detected: Bifurcation 1 at $\kappa = 0.4592$, Bifurcation 2 at $\kappa = 0.6429$, and Bifurcation 3 at $\kappa = 0.7959$.

The visualization includes six complementary plots: period versus $\kappa$, which detects divergences; maximum Lyapunov exponent versus $\kappa$; bifurcation indicator ($1/\text{period}$) versus $\kappa$, which highlights homoclinic bifurcations; Poincar\'e sections for sample $\kappa$ values; a 3D trajectory at a bifurcation point, showing the homoclinic structure; and the distribution of $x_2$ on the Poincar\'e section for different $\kappa$, providing a statistical bifurcation signature.

\begin{figure}[H]
\begin{center}
 \begin{minipage}{7.1in}
\includegraphics[width=7.1in]{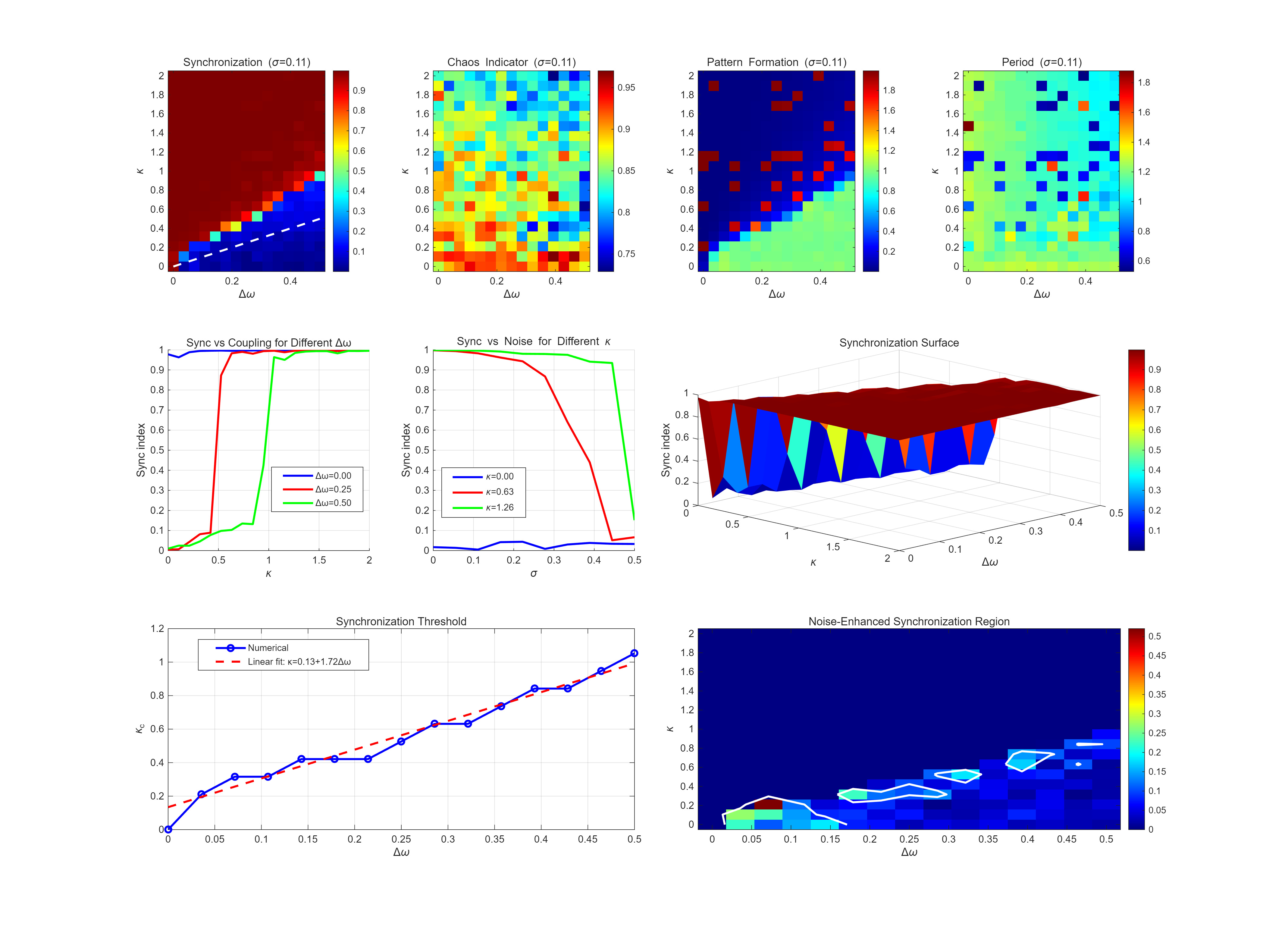}
\end{minipage}
\caption{We produce several plots to visualize the phase diagrams in the ($\kappa$, $\Delta\omega$) plane for a fixed noise level, as well as cross-sections and 3D surfaces. We fit a theoretical curve for the synchronization threshold and identify regions of noise-enhanced synchronization.}\label{Code10}
\end{center}
\end{figure}

Fig. \ref{Code10}  presents the results of a comprehensive parameter space exploration.
We explore a 3D parameter space ($\kappa$, $\Delta\omega$, $\sigma$) for two coupled van der Pol oscillators with noise in \eqref{CSVdP}. For each parameter set, we compute four metrics: synchronization index, chaos indicator, pattern formation index, and mean period.

We loop over a grid of parameters: coupling strength $\kappa$, frequency mismatch $\Delta\omega$, and noise intensity $\sigma$. For each parameter set, we simulate the two coupled oscillators and compute the metrics.
The metrics are defined as follows: the synchronization index, given by the phase coherence $R$ of the two oscillators (with $R = 1$ indicating perfect synchronization); the chaos indicator, an approximate Lyapunov exponent (where positive values signify chaos); the pattern formation index, calculated as the variance of the difference between the two oscillators normalized by the total variance (with high values indicating pattern formation, i.e., the oscillators exhibit different behaviors); and the mean period, estimated from the zero crossings of $x_1$.

We fix $\mu=0.5$ (nonlinearity parameter), corresponding to weakly nonlinear oscillators, while varying the other parameters. The visualization consists of 12 subplots. These include 2D heatmaps of the metrics in the ($\kappa$, $\Delta\omega$) plane for a fixed noise level; cross-sections showing how the synchronization index varies with $\kappa$ for different $\Delta\omega$, and with $\sigma$ for different $\kappa$; a 3D surface of the synchronization index; a plot of the synchronization threshold ($\kappa_c$ vs. $\Delta\omega$) with a linear fit; and a heatmap of noise-enhanced synchronization, highlighting regions where adding noise improves synchronization.

For small $\Delta\omega$ and large $\kappa$, the oscillators synchronize. The Arnold tongue is typically V-shaped, meaning the critical coupling $\kappa_c$ increases with $\Delta\omega$. A white dashed line for the theoretical boundary ($\kappa = \Delta\omega$) provides a simple approximation. The heatmap of the synchronization index should show high values (close to 1) for large $\kappa$ and small $\Delta\omega$, with the boundary resembling an Arnold tongue. The key observation is that synchronization requires $\kappa > 2.652\Delta\omega$.

\section{Large Networks}\label{LN}
The continuum limit of a large network of coupled van der Pol oscillators is given by
\begin{equation}\label{KM}
 \frac{\partial^2 X}{\partial t^2}-\mu(1-X^{2})\frac{\partial X}{\partial t}+\omega_0^{2}X=D\frac{\partial^2 X}{\partial \theta^2}+\sigma\xi(\theta,t),
\end{equation}
where $\theta$ is the spatial coordinate (e.g., around a ring), $D$ is the diffusion constant, and $\xi(\theta,t)$ is spatiotemporal white noise.

In the deterministic case ($\sigma=0$), linear stability analysis assuming $X(\theta,t)= e^{\lambda t+ik\theta}$ yields the dispersion relation
\begin{equation*}
\lambda^{2}-\mu\lambda+(\omega_{0}^{2}+D k^2)=0,
\end{equation*}
and pattern formation occurs for $D < 0$ (anti-diffusion).

The equation \eqref{KM} exhibits noise-induced patterns (for $\sigma>0$) and collective chaos, where noise excites spatial modes absent in the deterministic system (even with
$D>0$),  as quantified by the structure factor
\begin{equation*}
S(k,\omega)=\frac{\sigma^2}{|\lambda(k,\omega)|^2},
\end{equation*}
where $\lambda(k,\omega)$ is the eigenvalue of the linear operator.  This also captures noise-induced Turing patterns when diffusion coefficients depend on amplitude. Meanwhile, the deterministic system in the absence of noise may display spatiotemporal chaos, which noise can modify or induce in forms such as amplitude chaos (in $|X(\theta,t)|$), phase chaos (in $\phi(\theta,t)$), or defect-mediated turbulence (via chaotic motion of phase singularities).

The Lyapunov spectrum for the stochastic partial differential equation \eqref{KM} scales as
\begin{equation*}
\lambda_j\sim-\frac{D}{l_c^{2}}j^2\quad (\text{for large} j),
\end{equation*}
where $l_c$ is  the correlation length.

Synchronization transitions in networks are characterized by the order parameter
\begin{equation*}
Z(t)=\frac{1}{L}\int_{0}^{L}e^{i\phi(\theta,t)}d\theta,
\end{equation*}
and in the  $(\sigma,D,\mu)$ parameter space, the phase diagram includes coherent states ($|Z|\approx 1$), incoherent states ($|Z|\approx 0$), chimera states featuring coexisting coherent and incoherent domains, and turbulent states exhibiting spatiotemporal chaos.

\begin{figure}[H]
\begin{center}
 \begin{minipage}{6.4in}
\includegraphics[width=6.4in]{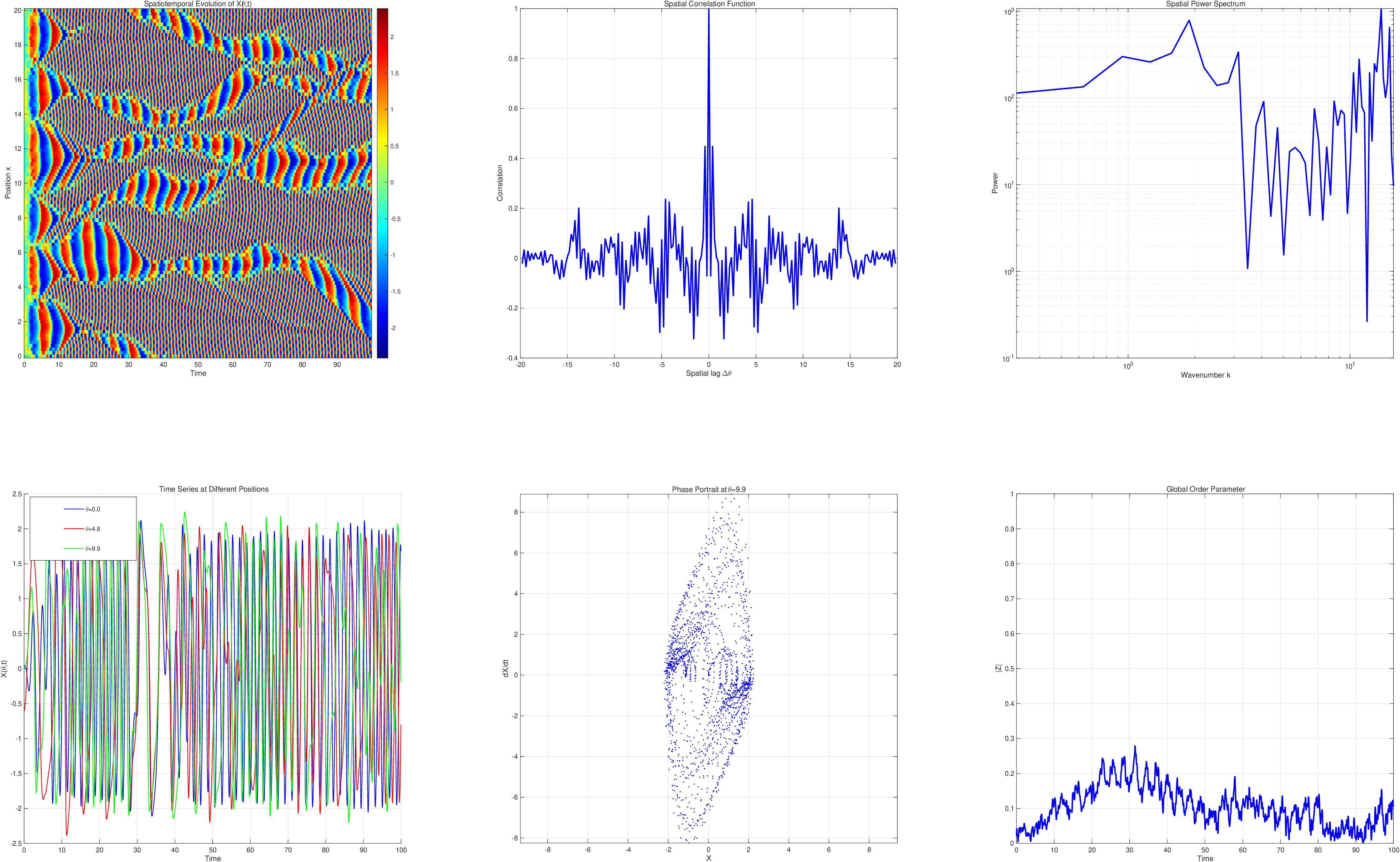}
\end{minipage}
\caption{For the Kuratani model \eqref{KM} in the continuum limit with noise, we sketch the spatiotemporal evolution, spatial correlation, spatial power spectrum, time series at different points, phase portrait, and global order parameter.}\label{Code61}
\end{center}
\end{figure}
We produce two figures with multiple panels.
The parameters are chosen as follows: $\mu=2$ (nonlinear damping),
$\omega_0=1$ (natural frequency), $D=0.1$ (diffusion coefficient, which is positive and thus stabilizing), $\sigma=0.3$ (noise intensity), $L=20$ (domain length), $N_x=100$ (spatial points, giving $dx=0.2$), $dt=0.05$, and  $T=100$ (so $N_t=2000$).
The initial condition is a sine wave with wavenumber 3 (i.e., three full waves in the domain) plus small random noise.

The Laplacian is discretized using a central difference scheme with periodic boundary conditions, implemented by constructing a sparse matrix $\mathbf{D}_2$.
The integration scheme is Euler-Maruyama for the stochastic partial differential equation \eqref{KM}:
\begin{align*}
Y_\text{new}&=Y_\text{old}+[\mu(1-X^2)Y-\omega_0^2X + D(\Delta X)]dt+\sigma dW,\\
X_\text{new}&=X_\text{old}+Y_\text{new}dt,
\end{align*}
which is a first-order scheme. Note that noise is added only to the velocity equation.
The noise is generated as a matrix of independent Gaussian random variables for each spatial point and time step; it is additive and white in both space and time.

Fig. \ref{Code61} presents an analysis of pattern formation in the Kuratani model, which is a continuum limit of coupled Van der Pol oscillators with noise. We simulate a stochastic partial differential equation \eqref{KM} and analyze pattern formation. The estimated pattern wavelength from autocorrelation peaks is 1.255, with 12 peaks found at lags 2, 14, 18, 21, 23, 32, 34, 37, 39, 48, 69, 71. FFT-based wavelength estimation gives
 0.455, corresponding to a dominant wavenumber of 13.823.

Fig. \ref{Code61} contains six panels showing spatiotemporal evolution of $X$ (displayed as an imagesc plot), spatial correlation function of the final profile, spatial power spectrum (log-log plot of $|\text{FFT}(X)|^2$), time series at three fixed positions, phase portrait ($X$ vs $dX/dt$) at the middle point, and global order parameter (amplitude of the spatial average of $\exp(i\text{phase}))$.

\begin{figure}[H]
\begin{center}
 \begin{minipage}{6.5in}
\includegraphics[width=6.5in]{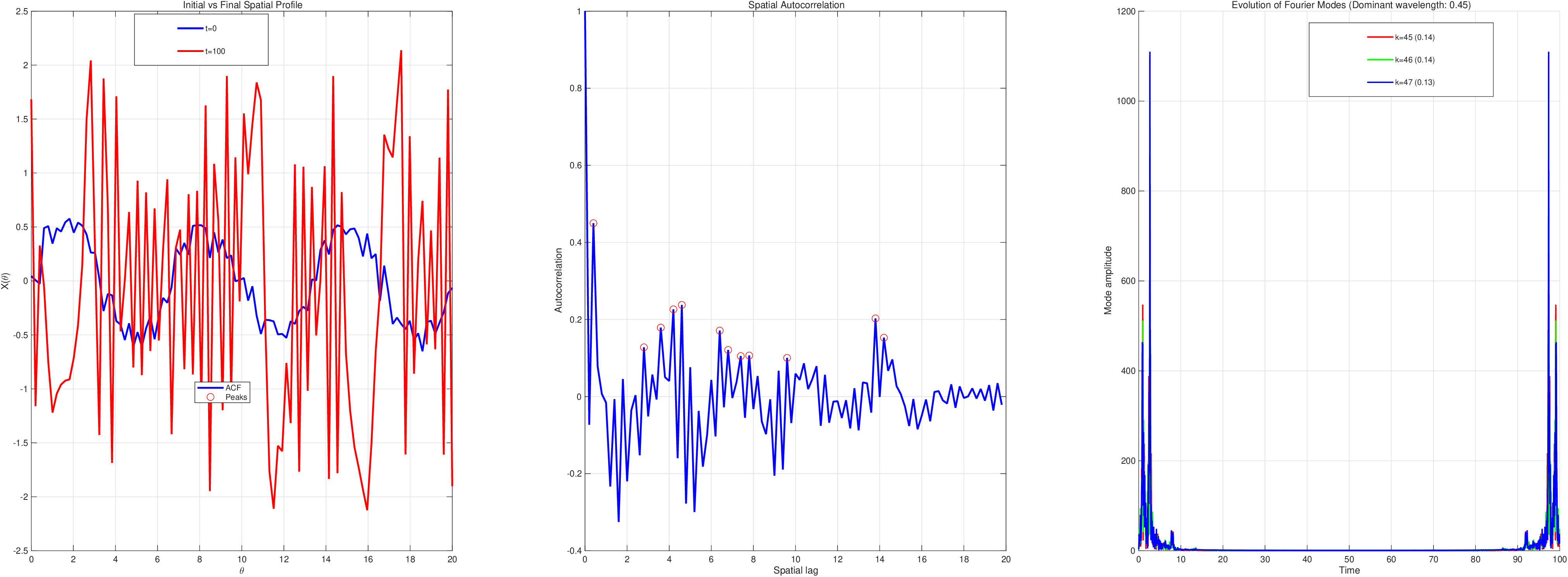}
\end{minipage}
\caption{For the Kuratani model \eqref{KM} in the continuum limit with noise, we plot the initial and final spatial profiles, the spatial autocorrelation (with peak finding for wavelength estimation), and the evolution of Fourier modes.}\label{Code62}
\end{center}
\end{figure}

Fig. \ref{Code62} contains three panels showing initial and final spatial profiles, spatial autocorrelation function (ACF) of the final profile with peak finding to estimate the pattern wavelength, and evolution of the amplitudes of the dominant Fourier modes over time.

\begin{figure}[H]
\begin{center}
 \begin{minipage}{6.2in}
\includegraphics[width=6.2in]{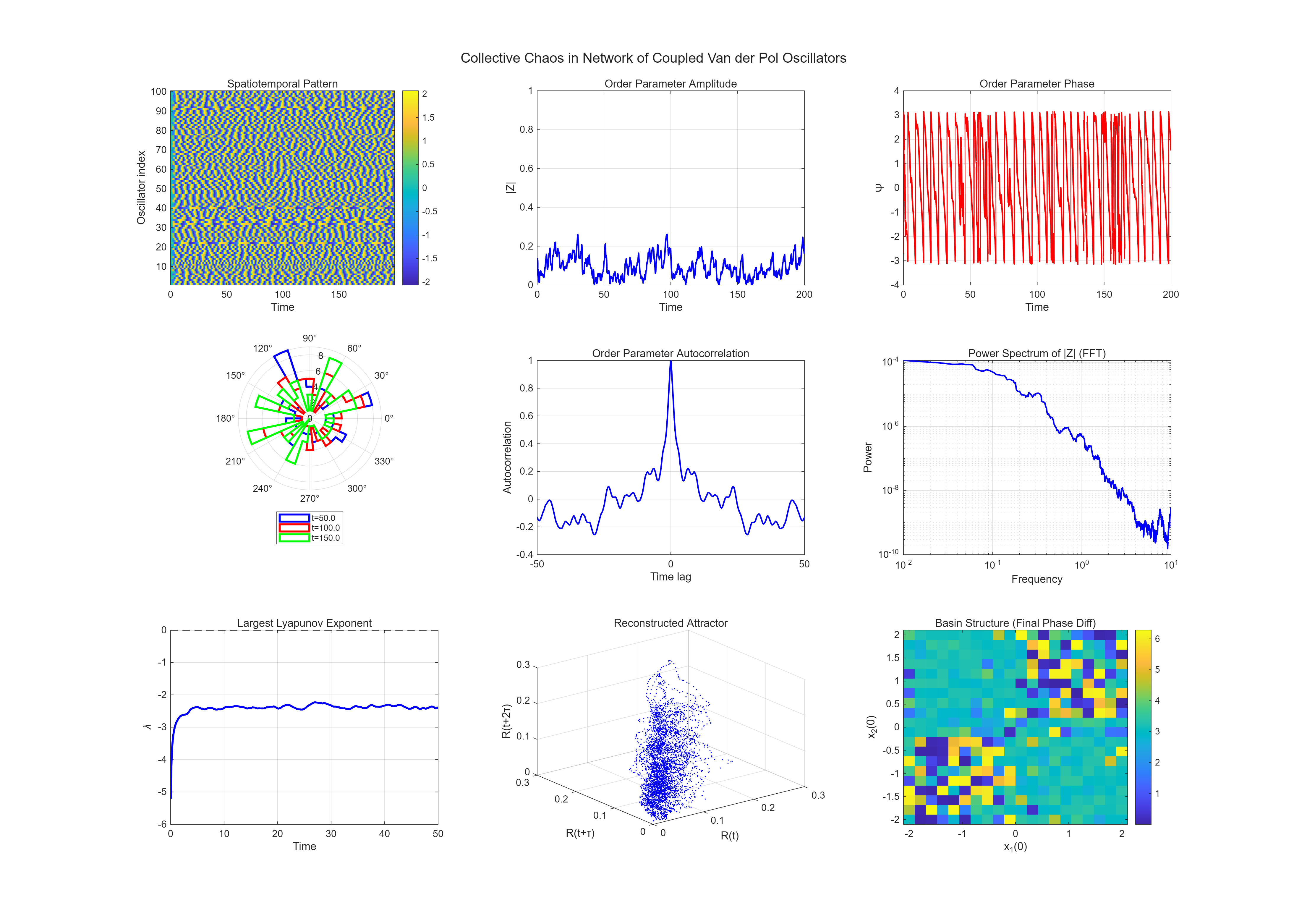}
\end{minipage}
\caption{Collective chaos and spatiotemporal dynamics.}\label{Code7}
\end{center}
\end{figure}

In Fig. \ref{Code7}, we simulate a network of coupled stochastic van der Pol oscillators and analyze collective chaos. We use global (mean-field) coupling, calculate the order parameter and its dynamics, and attempt to compute the largest Lyapunov exponent from the order parameter time series.

We investigate collective chaos in globally coupled stochastic van der Pol oscillators-a paradigmatic system for studying the emergence of complex dynamics from simple interacting units. The system is essentially the stochastic mean-field van der Pol model:
\begin{equation*}
\ddot{x}_{i}-\mu(1-x_{i}^{2})\dot{x}_{i}+\omega_{i}^{2}x_{i}=\kappa(\langle x\rangle-x_{i})+\sigma \xi_{i}(t),
\end{equation*}
where $\langle x\rangle=\frac{1}{N}\sum_{j}x_{j}$ is the mean field.

Global (mean-field) coupling implements all-to-all identical coupling with strength $\kappa$; it is equivalent to diffusive coupling with uniform weights. The natural frequencies $\omega_{i}$ are drawn from a Cauchy distribution with scale parameter
0.1, which has infinite variance (heavy tails) and is analytically tractable in mean-field theories. This choice introduces moderate heterogeneity.

Expected order parameter dynamics (for parameters $\kappa=0.3$, $\sigma=0.2$, frequency width $0.1$): From the Kuramoto model analogy, the theoretical critical coupling is $\kappa_c\approx0.1$.
With $\kappa=0.3>0.1$, we expect partial synchronization. Noise with intensity $\sigma=0.2$ reduces the synchronization level. Without noise, the order parameter
 $R$ would lie in the range $0.6-0.8$; with noise, we expect
 $R\approx0.4-0.6$. The instantaneous order parameter $R(t)$ exhibits temporal fluctuations.

We use time-delay embedding with embedding dimension $m=3$ (reasonable for a low-dimensional attractor) and delay
$\tau=10$ ((corresponding to  $10dt=0.5$ time units).

For globally coupled oscillators, several types of collective chaos can appear: amplitude-mediated collective chaos, where macroscopic amplitude fluctuations are chaotic while individual oscillators remain partially synchronized; phase turbulence, in which the phase field becomes turbulent and $|Z|$ fluctuates but remains positive; and intermittent collective chaos, characterized by bursts of chaos interspersed with regular dynamics.

For the Kuramoto model with inertia (similar to the van der Pol oscillator):
\begin{equation*}
I\frac{d^{2}\theta_i}{dt^{2}}=-\frac{d\theta_i}{dt}+\omega_{i}+\kappa R\sin(\psi-\theta_{i})+\xi.
\end{equation*}
the mean-field equations take the approximate form
\begin{align*}
\frac{dR}{dt}&=-R+\kappa R(1-R^2)/2+\xi,\\
\frac{d\psi}{dt}&=\Omega.
\end{align*}
and chaos can emerge through a Hopf bifurcation of these mean-field equations.

For the chosen parameters ($\kappa=0.3$, $\sigma=0.2$, frequency width $0.1$), the expected behavior is as follows: without noise, the system exhibits either periodic or quasiperiodic collective dynamics, whereas with noise ($\sigma=0.2$), the noise may induce or enhance chaos, yielding a small positive or near-zero Lyapunov exponent $\lambda$.

Finite-size effects are also present. With $N=100$ oscillators, the simulation captures moderate finite-size effects. A proper analysis would require examining how the results depend on $N$ (finite-size scaling) and how they approach the thermodynamic limit $N\rightarrow\infty$, where the behavior may differ from finite-$N$ simulations.
We simulate the model \eqref{KM} for pattern formation. The model is a spatially extended version of the van der Pol oscillator with diffusion and noise. We explore how noise can induce patterns that resemble Turing patterns, even in the absence of a deterministic Turing instability.

\begin{figure}[H]
\begin{center}
 \begin{minipage}{7.1in}
\includegraphics[width=7.1in]{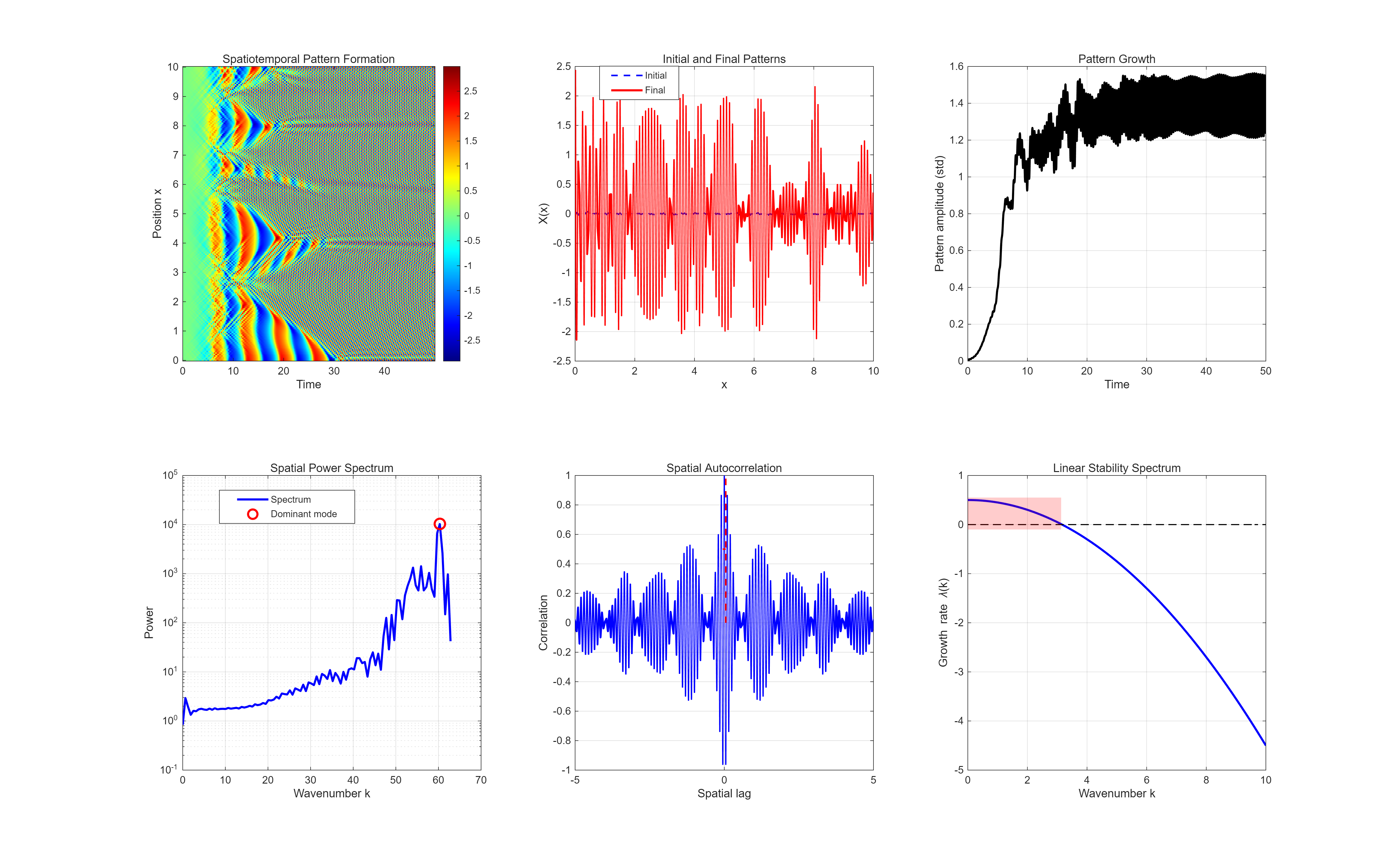}
\end{minipage}
\caption{The plot includes spatiotemporal pattern visualization, the evolution of pattern amplitude (standard deviation in space) over time, Fourier analysis to identify the dominant wavelength, spatial autocorrelation to estimate the correlation length, and linear stability analysis of the homogeneous state.}\label{Code81}
\end{center}
\end{figure}

Fig. \ref{Code81} presents an analysis of noise-induced pattern formation in a stochastic partial differential equation \eqref{KM}. The dominant wavelength estimated from the pattern is 0.104, the correlation length is 0.05, and the range of unstable wavenumbers extends from 0 to 3.131.

The parameter values are set as follows: the nonlinear damping parameter is $\mu=1.5$, the natural frequency is $\omega_0=1$, the diffusion coefficient is $D=0.05$, and the noise intensity is $\sigma=0.1$. The domain length is $L=10$ with $N_x=200$ spatial points, giving a spatial resolution of $dx=0.05$. The simulation uses a time step of $dt=0.02$ and runs for a total time $T=50$, corresponding to $N_t=2500$ time steps.

Spatial discretization uses Neumann boundary conditions. We generate both white and colored noise; the colored noise is spatially correlated using a Gaussian kernel with a specified correlation length. The integration employs the Euler-Maruyama method for the stochastic differential equations. The system consists of a second-order oscillator at each spatial point, coupled through diffusion.

We simulate the following system:
\begin{align*}
\dot{X}&=Y,\\
\dot{Y}&=\mu(1-X^{2})Y-\omega_0^{2}X+D\nabla^{2}X+\sigma\xi(x,t),
\end{align*}
where $\xi(x,t)$ is the noise (white or colored in space).
The variable $X$ represents the displacement, and $Y$ is the velocity. The diffusion term acts only on $X$
(second derivative in space), and the noise is added only to the $Y$ equation, which is typical for a driven oscillator.

The deterministic part (without noise and diffusion) is a van der Pol oscillator. The diffusion term couples the oscillators in space. The parameter $D=0.05$ is positive, which usually leads to dissipation and stabilization. The noise is additive and can be spatially correlated.

We investigate the emergence of patterns from a homogeneous state (with small random perturbations) under the influence of noise. The patterns are analyzed in terms of their amplitude, dominant wavelength, and correlation length.

A linear stability analysis of the homogeneous steady state ($X=0$, $Y=0$) is performed. The linearized equations (without noise) are
\begin{align*}
\dot{X}&=Y,\\
\dot{Y}&=\mu Y-\omega_{0}^{2} X+D\nabla^2X.
\end{align*}
Assuming solutions of the form $e^{\lambda t+ikx}$, we obtain the characteristic equation
\begin{equation*}
\lambda^2-\mu\lambda+(\omega_{0}^{2}+Dk^2)=0.
\end{equation*}
The growth rate is the real part of $\lambda$, given by
\begin{equation*}
\lambda=\frac{\mu\pm\sqrt{\mu^2-4(\omega_{0}^{2}+Dk^2)}}{2}.
\end{equation*}
For the van der Pol oscillator with $\mu > 0$, the origin is unstable. In the presence of diffusion, the system may form patterns. The range of unstable wavenumbers (where $\operatorname{Re}(\lambda) > 0$) is from $0$ to $k_{\text{max}} \approx 3.131$ for the chosen parameters.

The analysis includes spatiotemporal pattern visualization, pattern amplitude (standard deviation in space) over time, fourier analysis to identify the dominant wavelength,
spatial autocorrelation to estimate the correlation length, linear stability analysis of the homogeneous state, and comparison of patterns induced by noise with different correlation lengths.

The spatiotemporal evolution reveals the emergence of structured patterns. The pattern amplitude grows and eventually saturates. The Fourier spectrum exhibits a dominant mode, indicating a pattern with a characteristic wavelength. The spatial autocorrelation function provides an estimate of the correlation length. The linear stability analysis shows a range of unstable wavenumbers (if the expression $\mu-\omega_{0}^{2}-Dk^2>0$).

The van der Pol oscillator has an unstable origin for $\mu > 0$, so the homogeneous state is unstable. In the presence of diffusion, the system forms patterns. These patterns are induced by noise, and their characteristics (such as the dominant wavelength) are influenced by the noise correlation length.

This model differs from standard Turing pattern models, which typically require two different diffusion coefficients for an activator and an inhibitor. Here, only one diffusion coefficient is present. Nevertheless, the noise can induce patterns that are Turing-like in appearance. The comparison of patterns for different noise correlation lengths demonstrates that the noise color can influence pattern selection.

\begin{figure}[H]
\begin{center}
 \begin{minipage}{6.1in}
\includegraphics[width=6.1in]{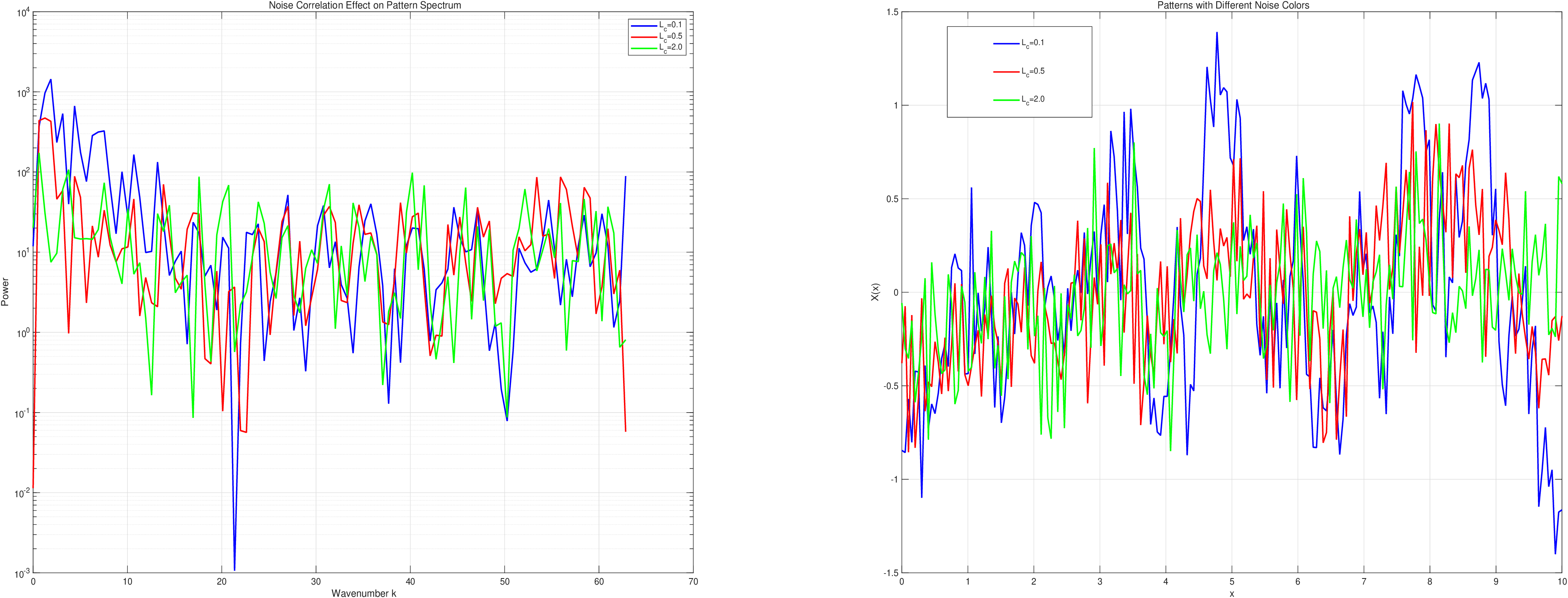}
\end{minipage}
\caption{Noise-induced patterns (Turing-like) in the Kuratani model \eqref{KM}.}\label{Code82}
\end{center}
\end{figure}

We perform two sets of simulations in Fig. \ref{Code82}: first with a fixed noise correlation length, and then a comparison of patterns for different noise correlation lengths. The results demonstrate how spatially correlated noise can induce patterns in a system of coupled oscillators, with the noise correlation length selecting the pattern scale.

\section{Conclusions and Future Challenges}\label{CFC}
We analyze the system \eqref{SVdP} of coupled stochastic van der Pol oscillators, focusing on global bifurcations, synchronization, and chaotic dynamics. We investigate how noise correlation structures influence synchronization, quantifying this effect by measuring average phase coherence and assessing synchronization enhancement due to common noise. The key parameters for analysis include the coupling strength $\kappa$, which controls synchronization; the noise intensity $\sigma$, which can induce or suppress chaos; the frequency width, which governs heterogeneity; and the nonlinearity $\mu$, which determines the dynamics of individual oscillators.

Bifurcation scenarios include a homoclinic bifurcation to an antisymmetric saddle, where increasing $\kappa$ causes the in-phase limit cycle to become homoclinic; and symmetry-breaking bifurcations, where symmetry breaks, leading to asymmetric oscillations. We perform a comprehensive analysis using Poincar\'e maps, presenting a robust framework that integrates statistical methods with geometric visualization to identify global bifurcations in stochastic systems. The period may vary with parameters; divergence can signal a homoclinic bifurcation, typically near bifurcation points. Regions may exist where synchronization increases with noise, a phenomenon linked to stochastic resonance. Chaotic behavior emerges at intermediate coupling strengths. A chaos indicator (e.g., Lyapunov exponent) is positive in chaotic regimes, typically when coupling induces complex dynamics without full synchronization. Chaos appear near bifurcation boundaries. The phase coherence index quantifies synchronous behavior, being high in synchronized states and low in desynchronized states.
This framework enables the study of collective chaos. It offers comprehensive visualization and employs multiple methods (e.g., order parameter, Lyapunov exponent, power spectra), providing educational insights into collective behavior.

In real systems (e.g., electronic circuits, neural oscillators), key signatures near bifurcations include: period divergence (excessively long periods); critical slowing down (heightened sensitivity to perturbations); and intermittency (trajectories lingering near saddles). Applications span coupled lasers, neural networks, power grids, and chemical oscillators. We will investigate a comprehensive framework for studying noise-induced transitions, integrating theoretical, simulation, and experimental approaches. We will also demonstrate how weak noise triggers rare yet significant events (phase slips) that alter synchronization dynamics, with implications for physics, engineering, and biology.

Future challenges include extending our framework to non-Gaussian fluctuations (e.g., $\alpha$-stable L\'evy processes \cite{AT,YLZ,YZD}) and complex network architectures \cite{BPK,LYX}. Such investigations are ongoing and will be reported in future work.
\medskip

\noindent\textbf{DATA AVAILABILITY}

The numerical algorithms and source code that support the findings of this study are available from the corresponding author upon reasonable request.

\medskip
\noindent\textbf{ACKNOWLEDGMENTS}

This work was supported by the Guangdong Basic and Applied Basic Research Foundation
(Grant No. 2025A1515012560), the Guangdong Introduction Program (Grant No. 2023QN10X753) and National
Foreign Experts Program (Grant No. 111001819820258003).

\end{document}